\documentclass[preprint,aps,floats,amsmath,showpacs,amssymb]{revtex4}

\usepackage{graphicx}
\usepackage{times}

\def\ie{{\it i.e.\;}}
\def\eg{{\it e.g.\;}}
\def\GeV{{\;\rm GeV}}

\def\TeV{{\;\rm TeV}}

\def\hp{H^+}
\def\hm{H^-}

\def\lsim{\mathrel{\raise.3ex\hbox{$<$\kern-.75em\lower1ex\hbox{$\sim$}}}}
\def\gsim{\mathrel{\raise.3ex\hbox{$>$\kern-.75em\lower1ex\hbox{$\sim$}}}}

\begin{document}

\preprint{$
\begin{array}{l}
\mbox{ANL-HEP-PR-03-056}\\
\mbox{MADPH-03-1334}\\
\mbox{CERN-TH/2003-291}\\
[5mm]
\end{array}
$}
\vspace*{2cm}
\title{Associated Production of a Top Quark and a Charged Higgs Boson}

\author{Edmond L Berger$^1$\footnote{berger@anl.gov}, 
        Tao Han$^{2}$\footnote{than@pheno.physics.wisc.edu}, 
        Jing Jiang$^{1}$\footnote{jiangj@hep.anl.gov},
        Tilman Plehn$^3$\footnote{tilman.plehn@cern.ch}}
\affiliation{$^1$High Energy Physics Division, Argonne National
  Laboratory, Argonne, IL 60439} 
\affiliation{$^2$Department of Physics, University of Wisconsin, 
Madison, WI 53706}
\affiliation{$^3$CERN Theory Group, CH-1211 Geneva 23, Switzerland}\bigskip

\date{\today}

\begin{abstract} 
  We compute the inclusive and differential cross sections for the
  associated production of a top quark along with a charged Higgs
  boson at hadron colliders to next-to-leading order (NLO) in
  perturbative quantum chromodynamics (QCD) and in supersymmetric QCD.
  For small Higgs boson masses we include top quark pair production
  diagrams with subsequent top quark decay into a bottom quark and a
  charged Higgs boson. We compare the NLO differential cross sections
  obtained in the bottom parton picture with those for the
  gluon-initiated production process and find good agreement. The
  effects of supersymmetric loop contributions are explored.  Only the
  corrections to the Yukawa coupling are sizable in the potential
  discovery region at the CERN Large Hadron Collider (LHC). All
  expressions and numerical results are fully differential, permitting
  selections on the momenta of both the top quark and the charged
  Higgs boson.
\end{abstract}


\maketitle

\section{Introduction}

The elucidation of electroweak symmetry breaking is an important goal
of particle physics. In the standard model, one neutral scalar Higgs
boson is assumed to exist, and it is associated with the generation of
the masses of the electroweak gauge bosons and of the fermions. The
neutral Higgs boson has not yet been observed, and direct searches
place a lower limit of about $114\GeV$ on its mass~\cite{lep}.
Extensions of the standard model include the possibility of more Higgs
fields.  The minimal supresymmetric standard model (MSSM) requires two
doublets to give mass to up-type and down-type fermions and to cancel
anomalies.  The doublets yield five physical Higgs bosons: two neutral
CP-even states, a CP-odd state, and a pair of charged scalars.  At
lowest order in perturbation theory, the masses and couplings of these
states depend on two parameters which may be chosen as the
pseudoscalar mass $m_A$ and the ratio of the two vacuum-expectation
values $\tan\beta = v_2/v_1$. Comprehensive analyses have been
performed of the expected coverage of the $(\tan\beta, m_A)$ parameter
space at the CERN Large Hadron Collider (LHC)~\cite{ATLASCMS}. While
the observation of at least one of the two CP even Higgs bosons may
not pose a problem for the LHC~\cite{nolose}, it will be challenging
to distinguish it from its standard model counterpart over a large
fraction of the parameter space. For small values of $\tan\beta$ the
only viable channel in which to observe a heavy Higgs boson could be
the resonant production of the scalar $H$ with subsequent decay to $hh
\to b\bar{b}\gamma\gamma$~\cite{neutral_pairs}, where $b$ is a bottom
quark.  For large values of $\tan\beta$, the identification of a
charged Higgs boson would provide evidence for a Higgs sector beyond
the standard model, meaning at least two Higgs doublets, and possibly
a supersymmetric Higgs sector.\medskip

If the charged Higgs boson is lighter than the top quark $t$, there is
a good chance that it will be discovered via the decay channel $t\to
b\hp$ in $p\bar p$ collisions at the Tevatron collider at $2\TeV$, or
in $pp$ collisions at the LHC at $14\TeV$.  Searches in the Run~I data
samples by the CDF and D0 collaborations at the Tevatron~\cite{tev}
place significant bounds on the mass $m_H$ and $\tan\beta$.  If the
charged Higgs boson is heavier than the top quark, then its
observation at hadron colliders becomes more problematic. In
particular, there is no tree level coupling of a single charged Higgs
boson to gauge boson pairs, and the production of $H^\pm$ is
inaccessible in weak boson fusion.  The cross section for $\hp \hm$
pair production is likely to be too small, and the heavy quark
backgrounds may be too large for the observation of charged Higgs
boson pairs, unless additional supersymmetric particles enhance this
loop-induced rate~\cite{charged_pairs_dy,charged_pairs_gg}. The
situation is similar for the associated production of a charged Higgs
boson with a $W$ boson. The standard model leads to a fairly small
rate, but supersymmetric particle loops might enhance the rate
considerably~\cite{charged_hw,zhu_hollik}.\smallskip

The most promising search channel for a heavy $H^\pm$ is the
associated production of a top quark and the charged Higgs boson
$pp\to t\hm+X$ and $pp\to \bar t \hp+X$ via the intermediary of 
a bottom quark coupling~\cite{charged_th}. Throughout this paper 
we present results only for the $t\hm$ channel, unless stated 
otherwise. If both the $t\hm$ and $\bar t \hp$ channels are included 
the rates increase by a factor of two. Advanced detector simulation 
studies have been done for the decay
channels $\hm \to \bar{t}b$~\cite{dec_top} and $\hm \to \tau
\bar{\nu}$~\cite{dec_tau}. The advantage of this production mode is
that the Yukawa coupling to a top quark and a bottom quark is enhanced
by a power of $\tan\beta$ for large values of $\tan\beta$, as are the
bottom quark Yukawa couplings of the heavy neutral Higgs bosons.  

We consider the leading order partonic subprocess to be 
\begin{equation}
gb\to t \hm, 
    \label{eq:gb}
\end{equation}
with the initial state bottom quark taken as a constituent of an 
incident proton.  The set of next-to-leading order subprocesses 
includes partonic reactions such as $gg \to \bar{b}t\hm$.  
The total rate for the process $pp\to \bar{b}t\hm + X$ receives large
corrections from collinear logarithms, originating from the radiation
of a forward bottom quark jet~\cite{bottom1,old}.  These logarithmic
terms can be resummed to all orders in the strong coupling strength
$\alpha_s$, leading to the bottom parton picture~\cite{bottom1} with
an appropriate bottom quark factorization
scale~\cite{old,bottom2,scott}. This resummation of large collinear
logarithms is valid not only for charged Higgs boson production, but
it is generic as long as there is a large mass scale $M$ that provides
$\log(M/p_{T,b})$ behavior.  In our case $M=m_t+m_H$.  The comparison
of higher order predictions for total cross sections of neutral Higgs
boson production shows impressive agreement between gluon-initiated 
and bottom parton results~\cite{bbh_nlo,robert}, but the absence of a 
heavy scale in
neutral Higgs boson production in association with bottom quarks may
have an impact on some final state distributions.\smallskip

The $t H^{-}$ production cross section can be evaluated with or 
without integration over the phase space of the final state bottom 
quark, corresponding to whether an accompanying final state bottom quark 
is observed or ignored.  The term ``exclusive'' is generally used to refer 
to a situation in which the final state bottom quark is observed, and 
``inclusive'' is used to refer to the case in which the final state bottom 
quark is ignored.  
\smallskip

Calculations of the next-to-leading order (NLO) total cross section 
for $pp\to t\hm+X$ are available in perturbative quantum 
chromodynamics (QCD)~\cite{zhu,old} as well as in supersymmetric 
QCD~\cite{old,Gao:2002is}.  If we use the same choice of parameters, 
most importantly the same renormalization and factorization scales, 
and the same scheme for renormalization for the Yukawa couplings, our 
numerical results are in good agreement with those of Ref.~\cite{zhu}.  
We comment 
in greater detail in Sec.~\ref{sec:nlo}~C about the differences in 
scheme choice.  The inclusion of NLO contributions to the $gb$ 
initial state process merges the ``inclusive-type'' process $gb\to t\hm$ 
with the ``exclusive-type'' process $gg \to \bar{b}t\hm$, whose 
contribution appears as part of the set of NLO diagrams.  The
NLO contributions increase the reliability of the theoretical
predictions by reducing the renormalization and factorization scale
dependence of the total rate.\medskip

In this paper, we present fully differential NLO cross sections for
the process $gb\to t\hm$. The differential distributions are
desirable, as are predictions of expected correlations among the final
state observables, since selections on final state kinematic variables
must be made in experimental studies, for reasons of event acceptance
and background rejection. In Sec.~\ref{sec:nlo}, we outline the
two-cutoff phase-space slicing method which we adopt.  We then study 
the NLO production rates at the LHC and the Tevatron with the associated
theoretical uncertainties. We present typical kinematic distributions
and momentum correlations in Sec.~\ref{sec:distri}.  For searches in
the framework of the MSSM we examine the effects of leading and
sub-leading supersymmetric QCD corrections in Sec.~\ref{sec:susy}.
Conclusions are summarized in Sec.~\ref{sec:con}.

The results in this paper go beyond those of Refs.~\cite{old,zhu}
in several respects.  The calculation presented in 
Ref.~\cite{old} uses a one-cutoff method for NLO calculations.  The 
two-cutoff method in this paper permits a fully differential 
treatment of the final state particles.  In this paper, we show 
that the bottom parton approach is justified for {\em differential} 
cross sections as well as for total cross sections.  A more extensive 
discussion of cutoff dependences is 
presented in this paper including two-dimensional plots.  In our  
treatment of SUSY-QCD corrections in Sec.~\ref{sec:susy}, we include 
an evaluation for the Snowmass points and slopes (SPS) parameters and 
the impact of resummation of the $\Delta_b$ corrections.  
Going beyond Ref.~\cite{zhu}, we include SUSY-QCD corrections, and
an exploration of the interplay of bottom parton and gluonic 
contributions. The matching of cross sections for small charged Higgs 
boson masses is new.  Matching near the top decay threshold is of 
considerable interest for LHC experiments.  

\section{Next-to-leading Order QCD Corrections}
\label{sec:nlo}

Throughout the paper we use a running strong coupling $\alpha_s$,
a bottom quark Yukawa coupling $y_b$, and a top quark Yukawa coupling
$y_t$ consistent with the order of perturbation theory,~\ie
one--loop running for the leading order and two--loop running for the
next-to-leading order results. If not stated explicitly otherwise we
neglect the bottom quark mass $m_b$ in the phase space as well as in
the matrix elements, while naturally keeping it in the bottom quark
Yukawa coupling. The Yukawa coupling is normalized to
$m_b(m_b)=4.2\GeV$ and a bottom quark pole mass of $4.6\GeV$.
Moreover, we use the CTEQ5 parton densities~\cite{cteq5}.  We refer to
the $K$ factor defined by $K = \sigma_{\rm NLO}/\sigma_{\rm LO}$, the
ratio of the NLO cross section over the leading order cross section.
The default choices of the renormalization scale and the factorization
scale are taken to be proportional to the hard scale in the
process~\cite{old,bottom2}
\begin{equation}
\mu^0_R=M/2 \qquad \qquad \mu^0_F=M/5 \qquad \qquad \qquad \qquad (M =
m_H+m_t). 
\label{eq:central_scale}
\end{equation}
A detailed explanation of these choices may be found 
in Sec.~\ref{sec:nlo}~B. 
\begin{figure}[t]
    \begin{center}
    \includegraphics[width=7.0cm]{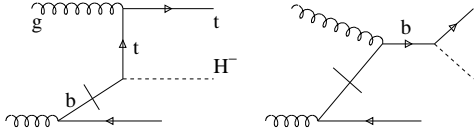}
    \end{center}
    \caption[]{The leading order Feynman diagrams for the production
    process $gb \to t \hm$. We indicate how the bottom partons are
    created through gluon splitting.}
    \label{fig:lofeyn}
\end{figure}

At leading order in QCD we start from the parton-level production
process in Eq.~(\ref{eq:gb}), with the diagrams depicted in
Fig.~\ref{fig:lofeyn}. It is appropriate to define the bottom quark as
a parton at high energies since the 
Dokshitzer-Gribov-Lipatov-Altarelli-Parisi (DGLAP) evolution
equation~\cite{Dokshitzer:1977sg} resums large logarithmic $\log
p_{T,b}$ contributions from small $p_{T,b}^{\rm min} \sim m_b$ to a
maximum value $p_{T,b}^{\rm max}$ (which in turn determines the
$b$-quark factorization scale). The bottom parton density is therefore
not suppressed by a simple power of $\alpha_s$. The leading order
cross section for a process involving one incoming bottom parton and
an incoming gluon is of order $\alpha_s y_{b,t}^2$, where $y_{b,t}$ is
the bottom-top-Higgs Yukawa coupling.  If $y_{b,t}$ is written as
$y_{b,t}^2 = y_t^2 + y_b^2$, $y_t$ and $y_b$ are terms proportional to
$(m_t/\tan\beta)$ and $(m_b \tan\beta)$ respectively.  
For large values of $\tan\beta$, $y_{b,t}$ is dominated by $y_b$.
The validity of our choice of the process in Eq.~(\ref{eq:gb}) as the
leading contribution is confirmed by our numerical results, namely
that the perturbative series is well behaved over a wide range of
scales.

There are two classes of NLO contributions:
\begin{itemize}
\item[(1)]
Virtual gluon exchange corrections to the lowest order process and the
corresponding real gluon emission corrections, both of order
$\alpha_s^2 y_{b,t}^2$,  
\begin{alignat}{7}
g b \;       &\to \; t H^- \qquad \qquad \text{(virtual correction)}
\nonumber \\
g b \;       &\to \; t H^- g.                                 
\label{eq:nlo_vproc}
\end{alignat}
\item[(2)]
The purely gluon-initiated and the purely quark-initiated diagrams, 
which lead to cross sections also of the order $\alpha_s^2 y_{b,t}^2$,
\begin{alignat}{7}
g g, \;  q \bar{q}, \;  b \bar{b}\;  &\to \; t H^- \bar{b},  \qquad\qquad
b      b \;   \to \; t H^- b,  \notag \\
b \bar{q} \; &  \to \; t H^- \bar{q}, \qquad \qquad 
b      q  \; \to \; t H^- q .
\label{eq:nlo_proc}
\end{alignat}
\end{itemize}
Because we neglect the bottom quark mass in the phase space and in the
matrix elements, the purely gluon and purely quark initiated
subprocesses are divergent in the collinear limit.  In our calculation
these divergences are removed through mass factorization, \ie the
proper definition of all parton densities at NLO.

One may think about an alternative treatment of the associated
production process, namely to start with the process $gg \to t\hm
\bar{b} $ as the leading contribution. These diagrams are
part of the $\alpha_s$ correction to the bottom--gluon
fusion process. For a choice of the factorization scales 
$\mu_{F,g} = \mu_{F,b} \to
m_b$ the bottom parton density vanishes, in contrast to the gluon
density, which is stable and well defined down to scales of the order
of $\Lambda_{\rm QCD}$. In a physical picture of this limit we consistently
switch off all large collinear logarithmic contributions, because
$p_{T,b}^{\rm max} \equiv \mu_{F,b}$. We are then
left with only the purely light-flavor $q\bar{q}$ and $gg$ induced
processes listed in Eq.~(\ref{eq:nlo_proc}). We use this limit of a
small bottom quark factorization scale in Sec.~\ref{sec:distri} to
check the impact of the bottom parton picture on the final-state
differential cross sections.

\subsection{Phase-space slicing}

One of the main tasks of our calculation is to integrate the
three-body matrix elements over the phase space of the unobserved
particle in the final state.  The situation is different from the case
of the single particle inclusive calculation in which one integrates 
over the phase space of two particles in the final state.  We wish 
to retain
control over the kinematic variables of a second particle in the
final state, while at the same time integrating over enough of the
phase space to ensure cancellation of all infrared and collinear
divergences.  Several techniques have been introduced for these
purposes. The phase-space slicing~\cite{slicing,brian} and the
subtraction methods~\cite{subtract} are two ways to extract the
singularities in the real emission contributions as exclusively as
possible.  All relevant information needed to compute a $2\to2$
particle NLO cross section with the two--cutoff slicing method is
compiled in Ref.~\cite{brian}. We follow this description closely in
our calculations.\medskip

The ultraviolet divergences in the virtual $2\to2$ corrections are
handled with dimensional regularization. The heavy final state masses
are renormalized in the on-shell scheme, while all couplings --- the
strong coupling as well as the bottom quark and the top quark Yukawa
couplings --- are renormalized in the $\overline{\rm MS}$ scheme. The
mismatch of the pre-factors between the virtual corrections and the
counter terms leads to the usual explicit $\log \mu_R$ dependence of
the NLO cross section on the renormalization scale.  The choice of 
the renormalization scale $\mu_R$ and that of the factorization 
scale $\mu_F$ are discussed in Sec.~\ref{sec:scale}.\medskip

Virtual gluon exchange and real parton emission lead to both soft
and collinear divergences.  They are extracted with dimensional 
regularization and partially canceled with each
other and partially removed through mass factorization, \ie the
consistent definition of parton densities. The situation is relatively
simple for processes that have different initial states from the
leading order $gb$ case because no soft divergence appears.
Schematically, we can write the contributions arising from the
processes in Eq.~(\ref{eq:nlo_proc}) as:
\begin{equation}
d \sigma_q = d \sigma_{2\to3,q}^{\rm HC} 
         + d \sigma_{2\to3,q}^{\rm finite} 
         + d \sigma_q^{\rm HMF}.
\end{equation}
The label HC indicates hard collinear divergences, which cancel with
the universal contributions from hard mass factorization (HMF). The
collinear phase space region, which appears for all $2\to3$
kinematics, is defined as the region in which the value of the
corresponding invariant for the two possibly collinear momenta $p_i$
and $p_j$ falls below $(p_i+p_j)^2 < \delta_c \; s$, where $\sqrt{s}$
is the partonic center of mass energy. The squared matrix element is
finite in the non-collinear phase space region and can be integrated
numerically, creating an implicit logarithmic dependence on the
cutoff $\delta_c$. Applying the hard mass factorization corrections,
we subtract the hard--collinear contributions as a convolution of the
leading order matrix element and the appropriate finite splitting
function, multiplied by $\log\delta_c$~\cite{brian}. The mismatch of
pre-factors leads to an additional explicit dependence of the NLO
cross section on the factorization scale $\log \mu_F$.\smallskip

The situation for virtual and real gluon emission in the $gb$
initial state is slightly more involved because additional
divergences appear due to soft gluon emission. Soft gluon emission is
defined by the non-invariant gluon energy constraint $E_g < \delta_s\;
\sqrt{s}/2$. The cross section can be written as 
\begin{equation}
d \sigma_g = d \sigma_{\rm virt}^{\rm S}
         + d \sigma_{\rm virt }^{\rm SC} 
         + d \sigma_{2\to3}^{\rm S} 
         + d \sigma_{2\to3}^{\rm SC} 
         + d \sigma^{\rm SMF} 
         + d \sigma_{2\to3,g}^{\rm HC} 
         + d \sigma_{2\to3,g}^{\rm finite} 
         + d \sigma_g^{\rm HMF},
\end{equation}
where the label S means soft, SC soft--collinear, and SMF 
soft--mass--factorization.  The additional soft $1/\epsilon$ and 
overlapping soft--collinear $1/\epsilon^2$ divergences appear in the 
virtual corrections as well as in the real gluon emission corrections.  
The divergences cancel among the virtual correction, the real gluon 
emission correction, and the contributions from mass factorization. 
We integrate numerically over the hard and non-collinear part of 
phase space and obtain an implicit dependence on
$\log \delta_s$ and $\log \delta_c$.\medskip

No explicit scale dependence occurs in dimensional regularization
after the purely soft divergences are canceled between the different
$d \sigma^{\rm S}$ and $d \sigma^{\rm SC}$ contributions, and the
pre-factors between real and the virtual gluon emission diagrams are
matched.  All poles in $d \sigma^{\rm S}$ and all double poles in $d
\sigma^{\rm SC}$ vanish after the soft divergences are removed, and
only single collinear poles $1/\epsilon$ remain in $d\sigma^{\rm SC}$.
Cancellation of these remaining divergences with the soft mass
factorization contribution renders a finite virtual gluon
emission matrix element with an additional explicit dependence on
$\log \delta_s$.  This logarithmic dependence cancels against the 
implicit dependence of the
numerically integrated hard non--collinear phase space. An explicit
dependence on the factorization scale $\log \mu_F$ remains in the
universal mass factorization terms.  Last, as a slight
complication, the same soft--collinear phase space configurations,
which include an explicit $\log \delta_c$ factor, also lead to an
implicit $\log \delta_s$ dependence after the numerical phase space
integration. This dependence again cancels with the cutoff dependence
of the $2\to3$ phase space, but it makes the numerical analysis
tedious~\cite{brian}.\smallskip

\begin{figure}[t]
 \includegraphics[width=15cm]{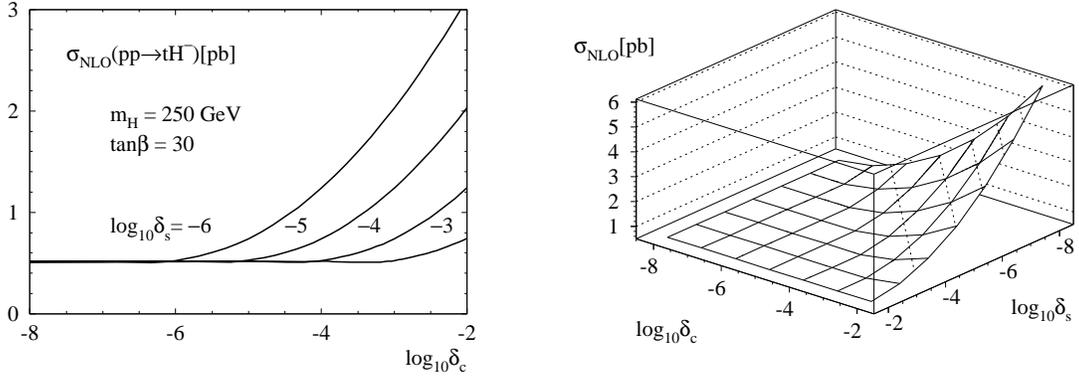}
 \caption[]{Cross section dependence on the cutoff parameters
     $\delta_c$ and $\delta_s$ at the LHC energy.  Left: The soft
     cutoff $\delta_s$ is fixed to different values. Right: two
     dimensional logarithmic dependence on both cutoff parameters.}
    \label{fig:cutoff}
\end{figure}

Both cutoff parameters, $\delta_s$ and $\delta_c$, must be small in
order that the soft and collinear approximation of the universal terms
be valid. Physical observables should not depend on the cutoff
parameters appreciably. We have checked in detail that the NLO cross
section and the distributions approach a two dimensional plateau for
sufficiently small values of $\delta_s$ and $\delta_c$. To define the
soft and the collinear regions of phase space consistently and to
avoid double counting, it is most convenient to require $\delta_c \ll
\delta_s$ and then determine the one dimensional plateau for a fixed
relation between $\delta_s$ and $\delta_c$.  We show the behavior of
the cross section versus the soft and the collinear cutoffs in
Fig.~\ref{fig:cutoff} at the LHC energy.  The cross section develops a
wide plateau for $\delta_c \lesssim \delta_s$. Unless noted otherwise
we use $\delta_c=10^{-5}$ and $\delta_s=10^{-3}$ in all numerical
analyses.  Checking the two dimensional plateau and comparing it with
results for the total cross section obtained with a one--cutoff
method~\cite{old}, we find that the NLO cross section has a remaining
uncertainty of $0.1\% - 0.5\%$ due to the cutoff dependence and the
corresponding numerical uncertainty.

\subsection{Scale Dependence}
\label{sec:scale}

Perturbative QCD calculations introduce an unwelcome dependence on the
renormalization scale $\mu_R$ and the factorization scale $\mu_F$.
One of the major motivations to perform NLO calculations is to reduce
this scale dependent theoretical uncertainty in predictions of
physical observables.  As the default renormalization scale, we choose
Eq.~(\ref{eq:central_scale}), related to the hard scale $M$.  We
identify the renormalization scales of the strong coupling and the
Yukawa coupling. This central renormalization scale choice leads to
perturbatively stable predictions for cross sections and branching
fractions as functions of $\alpha_s$~\cite{on_shell,scale_alphas} and
as functions of $y_b$~\cite{scale_mb}.  The situation is different for
the factorization scale. Two reasons point to a central scale
considerably smaller than the hard scale $M$, an optimum choice being
$\mu_F=M/5$.\smallskip

First, we can estimate the factorization scale from the kinematics of
the process $gg \to \bar{b}t\hm$. The bottom quark
factorization scale may be defined as the maximum $p_{T,b}$ which is
included in the $2 \to 2$ $gb \to t\hm$ process.  For the perturbatively 
calculated bottom quark density we can go back to the process $gg \to
\bar{b}t\hm$ and estimate up to which value $p_{T,b}^{\rm max}$ the
cross section shows the asymptotic behavior $\sigma_{\bar{b}t\hm} \sim
1/p_{T,b}$.  The hadronic phase space or, more specifically, the gluon
luminosity cuts off the asymptotic behavior near $p_{T,b}^{\rm max}
\sim M/5$.  This behavior can be understood independently from the
form of the matrix element, as long as both incoming partons are
either gluons or bottom quarks~\cite{old,bottom2}.  From basic
principles it is not clear if one could use the (factorizing) internal
momentum transfer $Q_b$ instead of $p_{T,b}$~\cite{scott}. The
difference in the maximum value up to which the asymptotic form holds
is $p_{T,b}^{\rm max} \sim Q_b^{\rm max}/2$, and the plateau in
$p_{T,b}$ is considerably softened~\cite{bottom2}.  We take this
uncertainty into account by varying the factorization scale over a
generous range.  Our argument works because the bottom parton density
is calculated perturbatively, meaning that its features are well
defined and understood perturbatively. \smallskip

Second, in the similar LHC process $b\bar{b} \to h$, the explicit
next-to-next-to-leading-order (NNLO) 
corrections are perturbatively most stable for, and therefore point
to, the same small factorization scale~\cite{robert}. Moreover, the
NLO corrections to the similar process $b\bar{b} \to W^+H^-$ are
negative for $\mu_F = M$~\cite{zhu_hollik}, indicating possibly a
collinear subtraction much too large.\medskip

\begin{figure}[t]
 \begin{center}
 \includegraphics[width=13.0cm]{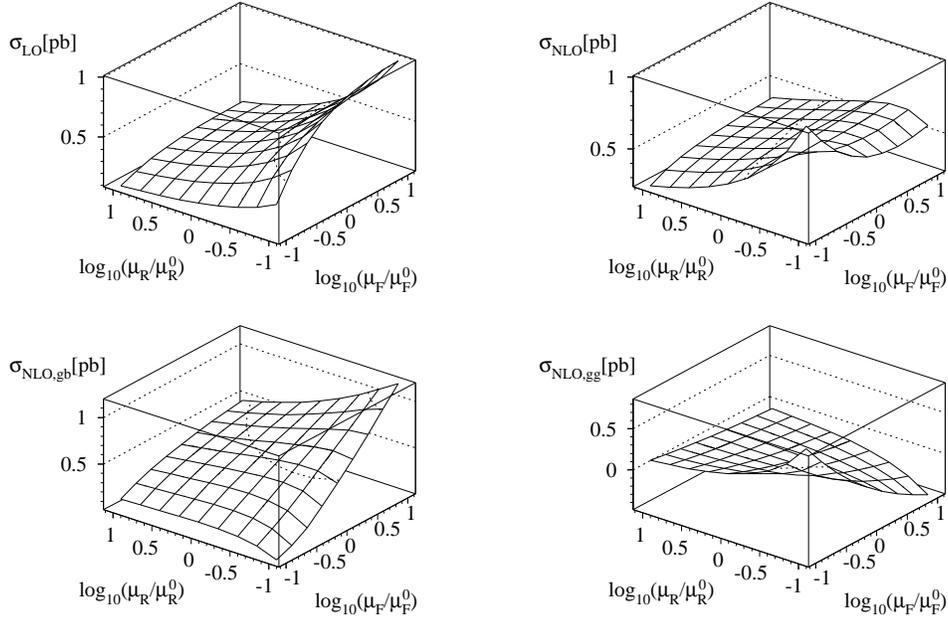}
 \end{center}
 \caption[]{The scale variation of the cross sections at the LHC for
   $m_H=250\GeV$ and $\tan\beta=30$. The central values are  
   $\mu_R^0=M/2$ and $\mu_F^0=M/5$. The four panels show the leading
   order cross section, the complete NLO cross section, the $gb$
   induced NLO cross section and the $gg$ induced contribution to the
   NLO cross section.}
 \label{fig:scale}
\end{figure}

The variation of the total cross section with the factorization and
renormalization scales is under control perturbatively if the two scales 
$\mu_R$ and $\mu_F$ are varied independently~\cite{old}.  However, there 
is a very large shift
in the total rate if the two scales are varied together $\mu_R \propto
\mu_F$ and run to very small values $\mu \lesssim M/10$. This behavior
suggests the presence of large contributions proportional to $\log
\mu_R \times \log \mu_F$. In Fig.~\ref{fig:scale} we show the scale
dependence of the different contributions to the leading order and
next-to-leading order cross sections for the process $gb \to t\hm$. 
The leading order curve (upper-left) behaves as
one would expect, namely the cross section increases for small $\mu_R$
and for large $\mu_F$, independently of each other. This behavior
arises from the running strong coupling $\alpha_s$ and the bottom
parton distribution.  In contrast, the running bottom Yukawa coupling
is relatively constant for these large scales. The NLO $gb$ initiated
curve (lower-left) is also easy to understand. The cross section
increases with the (bottom quark) factorization scale, but it develops
a maximum as a function of $\mu_R$ at a physical value of the scale,
for values of $\mu_F$ not too large.  The fraction $\sigma_{{\rm
  NLO},gb}/\sigma_{\rm LO}$ is under control perturbatively over the
range of scales, varying at most $20\%$. The corrections are largest
(and negative) for small $\mu_R$, because $\alpha_s$ is largest there.
Comparison of the NLO and LO results in the OS scheme with those 
obtained in the $\overline{\rm MS}$ scheme shows that use of the bottom 
quark pole mass as the Yukawa coupling produces perturbatively less 
stable rates, giving an unacceptably large leading order cross section.  
The $gg \to \bar{b}t\hm$ contribution
(lower-right) to the NLO rate has a very different behavior, which
dominates the complete NLO rate.  The $gg$ contribution is regulated
through mass factorization, meaning that we compute this process 
by subtracting out the contribution which is included in the
collinear bottom quark splitting.  For a central $\mu_R \sim \mu_R^0$,
the $gg$ channel gives a small positive contribution for small $\mu_F$
and a small negative contribution for large $\mu_F$. The latter
indicates that a choice of a large scale overestimates the logarithmic
terms, an overestimate then corrected by the explicit NLO diagram. If
we do not take into account the difference in size of the gluon and
bottom parton luminosities, the $gg$ initiated process is suppressed
by a factor $\alpha_s$ as compared to the $gb$ process.  The
pattern of correcting behavior stays but becomes much steeper if we
decrease $\mu_R$ and thereby increase $\alpha_s$.  For a central value
$\mu_F \sim \mu_F^0$ the $gg$ initial state corrections are zero, and
for much larger or smaller $\mu_F$ their absolute value increases
sharply. The $gg$ contribution grows for small $\mu_F$ and
simultaneously small $\mu_R$.  This effect can be rationalized if we
take the position that the choice of very small $\mu_F$ and very small
$\mu_R$ corresponds to a region in parameter space where the $gb$
initial state is not dominant perturbatively. The result arises
because we choose to ignore the presence of large collinear logarithms
up to $p_{T,b}^{\rm max} \equiv \mu_{F,b}$ and at the same time we
push $\alpha_s$ to large values.  The $K$ factor grows, indicative
that we should use the process $gg \to \bar{b}t\hm$ as the
leading process.

\subsection{Total cross section at the LHC}

\begin{figure}[t]
 \includegraphics[width=7.5cm]{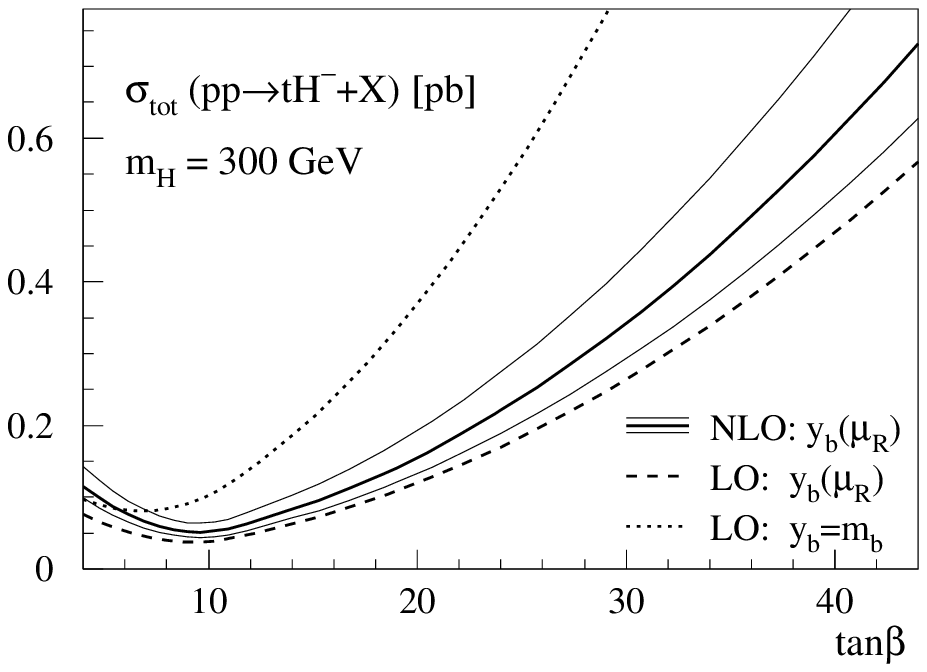}
 \hspace*{0.1cm}
 \includegraphics[width=8.1cm]{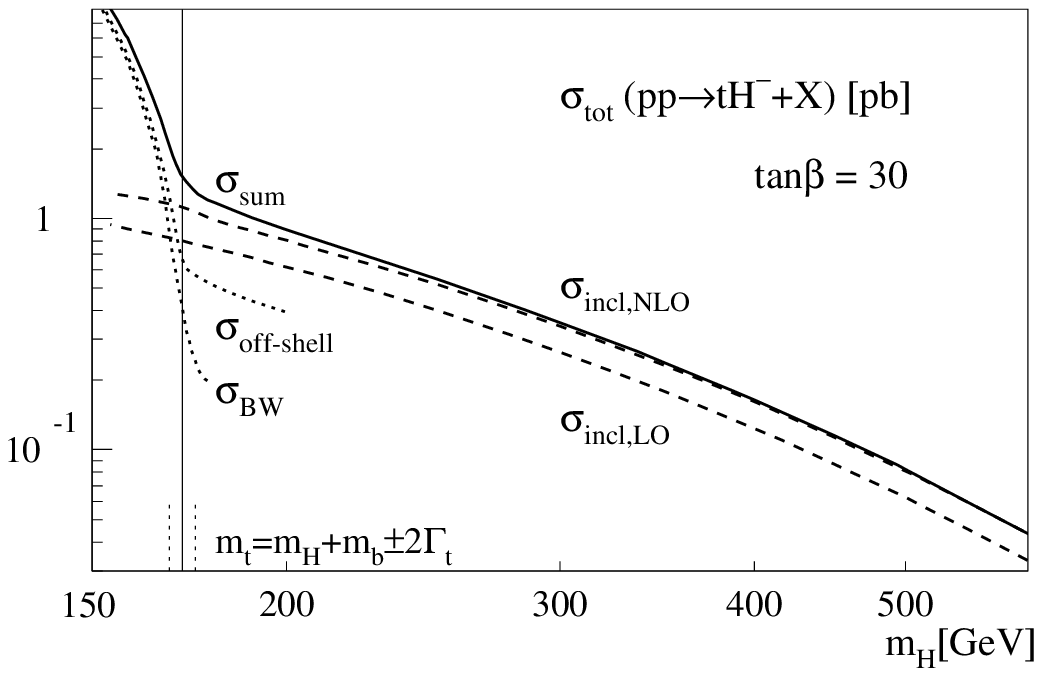}
 \caption[]{Left: The $t\hm$ inclusive cross section at the LHC
    as a function of $\tan\beta$. The solid curve is the NLO result,
    with the scale variations around the central scale
    $\mu/\mu_0=1/4-4$ indicated by the thinner solid curves.
    Also shown are the leading order result with a running bottom quark
    Yukawa coupling (dashed curve) as well as a pole mass Yukawa
    coupling (dotted curve).  Right: The total cross section at the
    LHC as a
    function of the charged Higgs boson mass as the solid curve.  The NLO
    and LO cross sections are shown as the dashed upper and lower
    curves, respectively.  The dotted curves show the cross sections for 
    $pp \to
    t \bar{t}^*+X$ with a subsequent decay $\bar{t}^* \to \bar{b} \hm$ in
    the Breit-Wigner approximation $(\sigma_{\rm BW})$ and that including
    the complete set of off-shell diagrams $(\sigma_{\rm off-shell})$.}
    \label{fig:tot}
\end{figure}

The effects of the NLO corrections on the total cross sections for the
process $gb \to t\hm$ at the LHC are shown in
Fig.~\ref{fig:tot}, versus $\tan\beta$ and versus the charged Higgs
boson mass $m_H$. The solid curve shows the NLO cross section, with
the scale variations around the central scale $\mu/\mu_0=1/4-4$
indicated by the thin solid lines in the left panel. The total cross
section increases for large values of $\tan\beta$, as expected for the
leading behavior $\sigma_{\rm tot} \propto \tan^2\beta$, while the $K$
factor is fairly independent of $\tan\beta$~\cite{old}. At leading
order we compare the cross section predictions for running bottom
quark Yukawa coupling and the pole-mass Yukawa coupling. The pole-mass
Yukawa coupling overestimates the total production rate, while the
leading order rate with a running bottom quark Yukawa coupling is a 
more reasonable approximation.  With the running bottom quark Yukawa 
coupling, the NLO result is slightly enhanced by a factor of $K \sim
1.4$~\cite{old}. The scale variation suggests a remaining theoretical
uncertainty of about $20\%$ on the predicted NLO cross section. Both
results confirm that the perturbative behavior of the production
process $gb \to t \hm$ is under control.

In the right panel of Fig.~\ref{fig:tot}, the NLO and LO cross
sections are indicated by the dashed upper and lower curves,
respectively.  The size of the NLO corrections is essentially
independent of the Higgs boson mass, a uniform enhancement factor of
about $1.4$. The process $gb \to t\hm$ itself is well defined over the
entire range of Higgs boson masses, as long as the hard scale
$M=m_H+m_t$ is sufficiently large to motivate the bottom parton
picture.

An interesting region is one in which the Higgs boson mass becomes
similar to or smaller than the top quark mass, and the decay $t \to b
\hm$ is possible. In Fig.~\ref{fig:tot} we see that the production of
a top quark pair $pp \to t\bar{t} + X$ with subsequent (off-shell)
decay $\bar{t}\to \bar{b}\hm$ becomes the dominant process.  For $m_H
\lesssim m_t$ the production cross section is of order $\alpha_s^2
y_{b,t}^2$ instead of $\alpha_s y_{b,t}^2$, and the large gluon
luminosity at the LHC is effective in this range of partonic energy.
In the following we discuss how these two processes can be combined,
to obtain a prediction of the cross section over the entire range of
Higgs boson masses.\medskip

For small Higgs boson masses, below the threshold $\bar{t} \to
\bar{b}\hm$, the $t\bar{t}$ production process with a subsequent decay
of the $\bar{t}$ dominates the rate for associated production of a
charged Higgs boson.  It is straightforward to combine the $t\bar{t}$
production process and the exclusive production channel $pp \to
\bar{b}t\hm+X$ with a tagged final state bottom-quark jet~\cite{nir}.
We compute the process $pp \to \bar{b}t\hm+X$ with a finite top-quark
width, essentially giving us a Breit--Wigner propagator for the
intermediate $\bar{t}$.  Approximate gauge invariance can be achieved
in the overall factor scheme~\cite{overall}, in which additional terms
${\cal O}(\Gamma_t/m_t)$ are traded for a gauge invariant matrix
element. Referring to the dotted curves in the right panel of
Fig.~\ref{fig:tot}, we see that the process $\bar{t} \to
\bar{b}\hm$ is approximated well by the subset of diagrams with an
intermediate $\bar{t}$ in the Breit--Wigner approximation, as long as
the Higgs boson mass is below the top quark threshold. Above the top
quark threshold, the exclusive process $pp \to \bar{b}t\hm+X$ is
dominated by the continuum off-shell diagrams, \ie the off-shell
extension of the $t\bar{t}$ production process, including all the
diagrams initiated from $q\bar q$ and $gg$.  These off-shell diagrams
become dominant where the cross section flattens and settles below the
$gb$ rate, while the Breit--Wigner cross section becomes very
small.\smallskip

For Higgs boson masses above threshold, a bottom quark jet tag is a
heavy price to pay, and it is likely a better idea to consider the
bottom--inclusive process. The collinear logarithms become large, and
to obtain the best possible prediction of the rate we must compute 
the process $gb \to t\hm$~\cite{old}, keeping in mind that
even at threshold the hard scale of the process is $M=350\GeV \gg
m_b$. The difference between the bottom--gluon induced rate
and the off-shell curve in Fig.~\ref{fig:tot} shows this enhancement
of the rate due to the resummed logarithmic terms.  Strictly
speaking, we should take into account that off-shell production also
includes the quark--initiated channels $q\bar{q} \to \bar{b} t \hm$,
which do not contribute to the bottom parton density at leading order.
However, they contribute only about $ 10\%$ to the total exclusive
rate.  The matching of the regions of small and large Higgs boson mass
in Fig.~\ref{fig:tot} indicates that a combination is needed of the
$t\bar{t}$ production process with the process $gb \to t\hm$. There 
appears to be no region of Higgs boson masses where
the off-shell production process $gg \to \bar{b}t\hm$ is the
appropriate perturbative description. \smallskip

In Fig.~\ref{fig:tot} we show the different $t\bar{t}$ cross sections
with a finite bottom quark mass, while we neglect the bottom quark
mass for the $gb$ channel. The uncertainty induced by this
approximation is small, however, since we cannot avoid neglecting
$\Gamma_t/m_t$ corrections and $m_b/M \sim \Gamma_t/M$.\medskip

At leading order the Breit--Wigner approximation of the $pp \to
t\bar{t}^* +X$ process, with subsequent decay of the off-shell
$\bar{t}^*$, may be
combined without problems with the process $gb \to t\hm$. We
may add the independent event samples for any Higgs boson mass value.
However, at NLO and for Higgs boson masses smaller than the top quark
mass there is a potential problem of double counting. The $t\bar{t}^*$
production process with a subsequent decay of the
$\bar{t}^*$ can be regarded as an ${\cal O}(\alpha_s)$ correction to the
$gb$ initiated process, Eq.~(\ref{eq:nlo_proc}), while it
can as well be viewed as (nearly) on-shell $t\bar{t}$ production with
a subsequent decay $\bar{t} \to \bar{b}\hm$.  To avoid double
counting, we subtract the resonant on-shell part of the $t\bar{t}$
diagrams from the NLO correction to $t\hm$ production and keep it
as part of the $pp\to t\bar{t} + X$ rate. The non-divergent off-shell
contribution of the $\bar{t}^*$ propagator is counted toward the NLO
$t\hm$ rate.  The division into on-shell and off-shell
contributions, however, is well defined only in the narrow width
approximation, so we neglect terms of order $\Gamma_t/m_t$. The
ambiguity reflects the unsolved problem of how a long-lived
intermediate particle is treated in field theory. Up to finite width
corrections we can, just as at leading order, add the rate for
$t\bar{t}$ production with a Breit--Wigner propagator and the properly
subtracted NLO $gb \to t\hm$ rate to obtain a prediction for any given
Higgs boson mass. In Fig.~\ref{fig:tot} we see that addition of the
cross sections is essentially equivalent to a naive matching
procedure. It is perhaps unexpected that the corrections from the
Breit--Wigner propagator extend to large Higgs boson masses. On the
other hand, since there is a $\sim 20\%$~\cite{old} theoretical
uncertainty on the NLO cross section for the $gb$ induced 
process, the details of this matching/adding procedure are not
important phenomenologically.  Instead of including the on-shell
production with the Breit--Wigner propagator for Higgs boson mass
values above $300\GeV$, we could as well have cut it off at $m_H
\simeq m_t+10\Gamma_t$.\smallskip
 
Below the $\bar{t} \to \bar{b} \hm$ threshold, the Breit--Wigner
description is valid, with higher order contributions included in the
cross section for $pp \to t\bar{t} +X$~\cite{Berger:1995xz,ttbar}.
Off-shell effects have little impact there. Above threshold these
off-shell effects have a considerable impact relative to the
Breit--Wigner description, but the $gb \to t\hm$ cross section is 
dominant. After the on-shell contributions are subtracted
from the NLO rate for $gb\to t\hm$, we can match the two results
simply by adding them, without any problem of double counting.

It is useful at this point to compare our final predictions with those
published in Ref.~\cite{zhu}.  For $m_H = 250$~GeV and $m_t = 175$~GeV,
and for the same value of $\tan\beta = 30$, our predicted $K$ factor is
$1.4$ versus about $1.6$ in Ref.\cite{zhu}.  However, these numbers should
not be compared directly since the factorization and the renormalization
scales are different in the two calculations.  In addition, there are
small differences in the values of parameters such as the NLO $m_b$, 
$\alpha_s$ and the choices of cutoffs.  If we use exactly the same 
parameters, cutoff choices, renormalization schemes, and most importantly 
the same factorization and renormalization scales, our result is 
$5$\% larger when compared to that of Ref.~\cite{zhu}.  We attribute this 
$5$\% difference to uncertainty in the numerical integration.  In the 
two-cutoff phase space method, the final NLO cross section is the difference 
of two large quantities.  The numerical uncertainty of either of the 
these integrations is less than $1\%$, but an uncertainty of $5\%$ can 
develop in the difference.

There are two masses in the matrix elements that must be renormalized,
$m_b$ and $m_t$.  The top mass $m_t$ enters both as an external quark mass
and in the Yukawa coupling.  The bottom mass $m_b$ appears only in the
Yukawa coupling, since we have set the external bottom quark mass to zero.
We consistently use the on-shell (OS) scheme for the external top quark
mass and we use the $\overline{\rm MS}$ scheme for the top and bottom quark
masses in the Yukawa couplings.  To understand the effect of a different
renormalization scheme for top quark mass in the Yukawa coupling, we
perform our calculation in both schemes.  The counter terms for the top
quark mass are
\begin{eqnarray}
\frac{\delta m_t^{\overline{\rm MS}}}{m_t} &=& - \frac{\alpha_s}{4\pi} C_F (4
\pi)^\epsilon \Gamma(1+\epsilon) \frac{3}{\epsilon_{UV}}~, \\
 \frac{\delta m_t^{OS}}{m_t} &=& - \frac{\alpha_s}{4\pi} C_F
\left( \frac{4\pi\mu^2}{m_t^2} \right)^\epsilon \Gamma(1+\epsilon)
\left(\frac{3}{\epsilon_{UV}} +4 \right)~,
\end{eqnarray}
for the $\overline{\rm MS}$ and the OS scheme respectively.  Changing from one 
scheme to another can induce about $12$\% difference in the NLO cross section
relative to the LO Born cross section.

We judge it physically more attractive to use the OS scheme 
for the external top quark mass since the top quark mass reconstructed 
in experiments is the pole mass.  Nevertheless, it is perhaps best to admit 
that the difference in scheme choice is tantamount to a difference at the 
next order in $\alpha_s$ and thus should be viewed as a theoretical 
systematic uncertainty at the order of perturbation theory in which 
we are working.

\subsection{Production at the Tevatron}

\begin{figure}[t]
 \begin{center}
 \includegraphics[width=8.5cm]{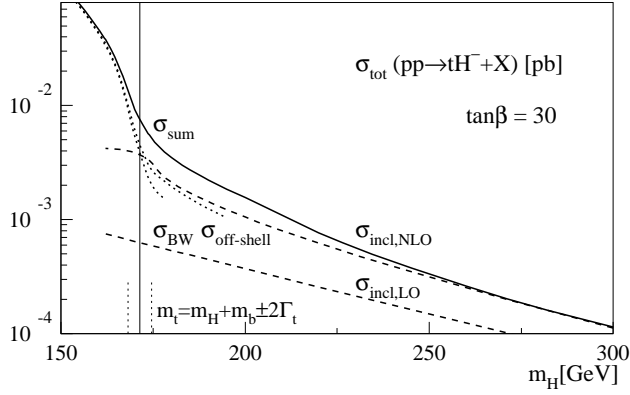}
 \end{center}
 \caption[]{The cross section for the Tevatron ($2\TeV$) as a function
    of the charged Higgs boson mass. We also show the cross sections for $pp
    \to t \bar{t}^* + X$ with a subsequent decay $\bar{t}^* \to \bar{b} \hm$
    in the Breit--Wigner approximation and including the complete set
    of off-shell diagrams.}
    \label{fig:tev}
\end{figure}

The successful matching of the Breit--Wigner approximation and 
the process $gb\to t\hm$ does not apply readily at the Tevatron.
In Fig.~\ref{fig:tev} we observe that the LO process $gb \to
t\hm$ underestimates the cross section compared to 
$\bar{b}t\hm$ production~\cite{nir}.  At the Tevatron the gluon
luminosity is not dominant in the relevant region of partonic
fractional momentum $x$. The gluon initial state contributes only
about $10\%$ to the total $p \bar{p} \to \bar{b}t\hm+X$ rate. Because 
initial state gluons are the dominant source of the bottom partons,  
we expect the leading order $gb$ rate to be far smaller than 
the true rate. The leading-order $2 \to 3$ processes contribute to 
the NLO $gb$ process, and perturbation theory for the 
$gb$ process is not well defined, in the sense that the NLO 
corrections are large.  The difference between the leading order $gb$  
rate and the off-shell $\bar{b}t\hm$ production rate is slightly less 
than a factor 10 because the $gb$ rate is still enhanced by the 
resummation of large logarithmic terms in the bottom parton picture.  
We limit ourselves to LHC results for most of the rest of this paper.

The NLO inclusive $t \hm$ rate consistently includes the whole set of 
quark initiated processes, Eq.~(\ref{eq:nlo_proc}). Because these quark
processes are dominant at Tevatron energies, the $K$ factor for the
$gb$ process can be as large as 5 for $m_H \sim 175\GeV$. The NLO
inclusive rate matches the exclusive rate from 
$p \bar{p} \to \bar{t}^* t +X$; $\bar{t}^* \to \bar{b} \hm$ fairly well,
particularly in view of possible remaining differences between these
two channels. The cross section $\sigma_{\rm off-shell}$  
for $p \bar{p} \to \bar{t}^* t +X$ is calcualted at leading order and  
evaluated with leading order running couplings and parton densities, 
while the quark-initiated contributions to the NLO inclusive rates 
$\sigma_{\rm incl,NLO}$ are evaluated
with NLO quantities.  Moreover, the collinear divergences in the
exclusive rate are regulated by a physical bottom quark mass, while
the NLO inclusive rate neglects the bottom quark mass and is regulated
by mass factorization, \ie by subtraction of the divergent
contributions to avoid double counting with the NLO evolution of the
parton densities.\smallskip

Just as for the LHC we see that the Breit--Wigner approximation and
the complete off-shell matrix element evaluation agree very well up to
$m_H \sim m_t$. Above threshold the LO $gb$ rate is significantly
smaller than the complete off-shell rate, but the NLO inclusive rate
matches the exclusive rate well.  The visible effect which the
Breit--Wigner contribution has on the matched/added sum of the cross
sections may be unexpected, but we keep in mind the substantial
theoretical uncertainty on the NLO prediction. The NLO
contribution from the $q\bar{q} \to \bar{b}t\hm$ production
process is larger than the LO $gb$ induced rate. The formally NLO $gb$
induced rate enters with a much wider band of uncertainty than the
$20\%$ we quote for the perturbatively well behaved LHC process. This
wider band covers different schemes for phasing out the Breit--Wigner
contribution toward large Higgs boson masses. In Fig.~\ref{fig:tev} we
cut off the Breit--Wigner cross sections for Higgs boson masses
between $200\GeV$ and $250\GeV$, \ie roughly 20 top quark widths above
threshold.\smallskip

Although we obtain a predicted cross section at the Tevatron for the
entire range of $m_H$, we emphasize that the matching of the
Breit--Wigner production process and $gb$ fusion works
only if we take into account the NLO corrections to the $gb$ 
channel. At the LHC the same matching of the two approaches at
threshold makes sense even for the leading order $gb \to
t\hm$ rate. Because the bottom parton picture is perturbatively stable
at the LHC energy, the prediction of the charged Higgs boson
production rate suffers from smaller theoretical uncertainty.

\section{Kinematic distributions}
\label{sec:distri} 

The bottom parton picture underlies the calculation of the production
cross section.  As summarized above, the bottom parton description
provides an appropriate way to compute the total cross section for
charged Higgs boson production if an appropriate bottom quark
factorization scale is used.  We establish three results in extending
our analysis to the kinematic distributions.  The normalized
distributions of sufficiently inclusive variables do not change
significantly from a LO to a NLO treatment of the process
$gb \to t\hm$.  Second, the distributions as well as the total rate do
not have a strong scale dependence. In particular, checking the limit
$\mu_F \to m_b$, we verify that the bottom parton picture does not
have much impact on the shape of the kinematic distributions of the
heavy final-state particles.  Finally, we test the approximation of
vanishing bottom quark mass in the phase space and the matrix
elements.\medskip

The first statement is easy to confirm.  We show the rapidity, the
transverse momentum, and the invariant mass distributions for the
heavy final-state particles in Fig.~\ref{fig:diff_nlo}.  Although the
additional bottom quark jet in the final state absorbs part of the
momentum from the incoming partons, the NLO transverse momentum
distributions are minimally harder.  The extra NLO purely
quark-initiated production process has a considerably harder $p_T$
spectrum, but it contributes only $10\%$ to the NLO rate (see also the
quark induced contribution in the left panel of
Fig.~\ref{fig:diff_mb}).  For single particle spectra at the LHC, we
conclude that the shift in the final state distributions from LO to
NLO is smaller than the typical scale uncertainty of $20\%$ on the NLO
rate. Similar behavior is found, for example, in the production of
heavy supersymmetric particles at
NLO~\cite{on_shell,Berger:1999mc}.\medskip

\subsection{Zero transverse momentum approximation}

\begin{figure}[t]
 \begin{center}
 \includegraphics[width=15.0cm]{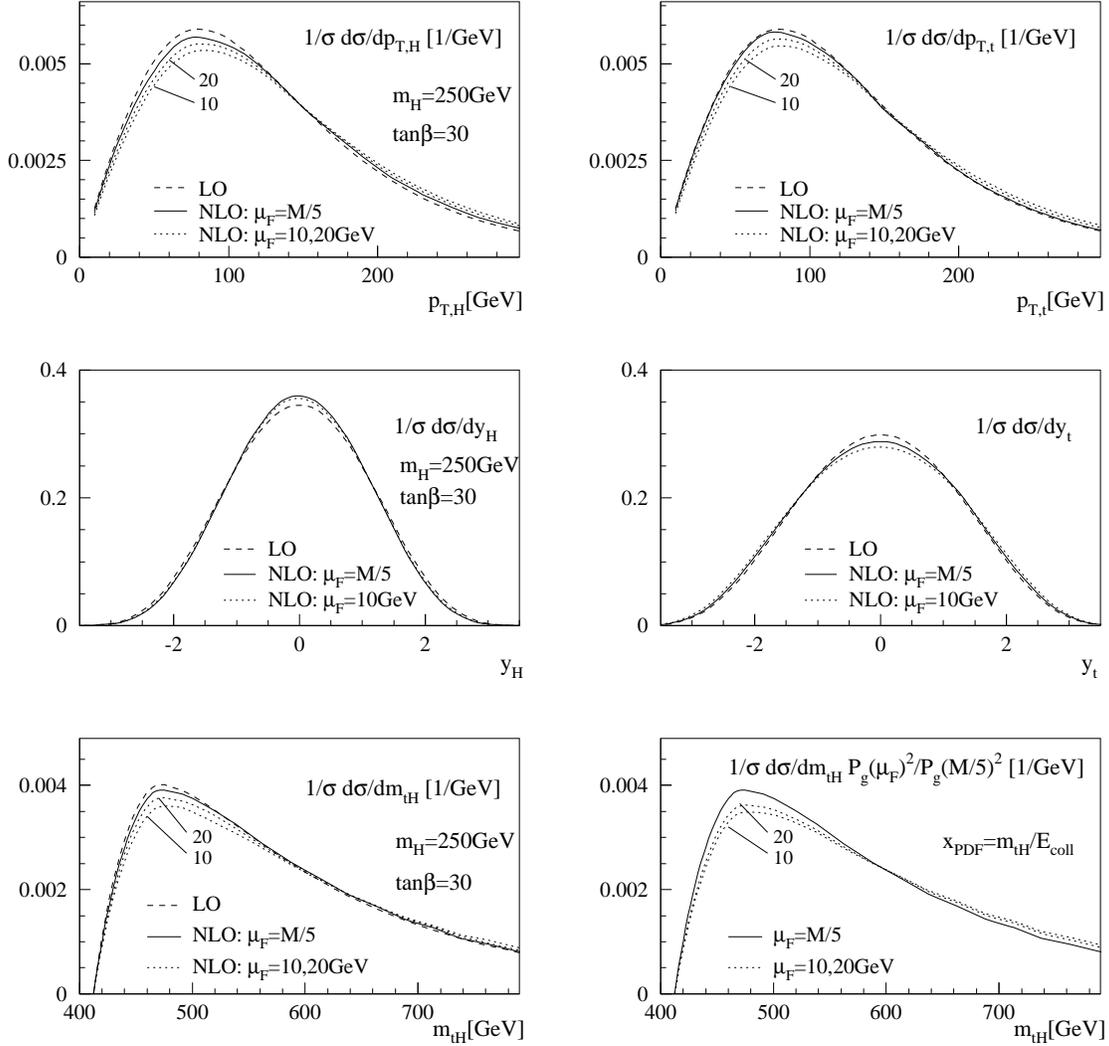}
 \end{center}
 \caption[]{The kinematic distributions for the heavy final-state
   particles at the LHC: transverse momentum and rapidity of the
   charged Higgs boson 
   (left) and the top quark (right) for the process $gb\to
   t\hm$. The dashed curve shows the leading order distribution with
   the central choice of scales. The solid and the dotted curves
   represent the NLO results for three choices of the bottom quark and
   the gluon factorization scale $\mu_F = M/5$, $\mu_F=20\GeV$ and
   $\mu_F=10\GeV$. The bottom row shows the distribution of the
   invariant mass of the top quark and the Higgs boson pair from the
   LO and NLO calculations, plus a rescaled NLO
   distribution (the last panel). The scaling factor involves the
   gluon parton densities at the different factorization scales.}   
 \label{fig:diff_nlo}
\end{figure}

The transverse momentum and the rapidity distributions of the heavy
final-state particles are depicted in Fig.~\ref{fig:diff_nlo}. The
choice of the bottom quark factorization scale $\mu_F = M/5$ has been
shown to be a part of a consistent bottom parton picture for this
class of processes at the LHC~\cite{bottom2,old,scott}. It remains to
be checked whether the collinear approximation for the gluon splitting
into a bottom parton is appropriate for the distributions of the final
state particles.\smallskip

We make use of the method described in Sec.~\ref{sec:scale} to test
the bottom parton approach and in particular the approximation of
negligible transverse momentum of the incoming bottom parton.  The
bottom parton density vanishes if the bottom quark factorization scale
approaches the bottom quark mass $\mu_{F,b}\to m_b$.  For consistency
reasons we use the same factorization scale for all partons, but
neither the gluon density nor the light quark densities change
dramatically as $\mu_F \to m_b \gg \Lambda_{\rm QCD}$. In this limit
the NLO $gb$ induced cross section at the LHC is dominated by the
process $gg \to \bar{b}t\hm$. Even though the final-state bottom quark
is massless in the calculation, the corresponding rate is finite and
well defined for any factorization scale $\mu_F > m_b$. All
divergences in the $gg \to \bar{b}t\hm$ process, regulated originally 
by a bottom
quark mass, are absorbed into the definition of the NLO parton
densities.  As described in Sec.~\ref{sec:scale}, when the bottom
quark factorization scale takes the limit $\mu_{F,b} \to m_b$ the
physical picture shifts from the resummed cross section, including a
large logarithmic term $\log p_{T,b}/m_b$, to the 
$\bar{b}tH^-$ situation in which the large logarithmic contribution
could be removed, for example, by a detector cut $p_{T,b}^{\rm min}$.
We show the difference between the NLO distributions with the central
factorization scale $\mu_F=M/5$ and the small scale limit in
Fig.~\ref{fig:diff_nlo}.  The scales we use are $\mu_F=10,20\GeV$.  In
principle, we could as well try $\mu_F=5\GeV$, for which the
calculated bottom quark density is well defined, but the parton
densities of gluons and light quarks are poorly constrained.  We
checked that we would then see all the features described below for
scales down to $10\GeV$, except that their effect on the cross
sections would be numerically more pronounced. \smallskip

We see in Fig.~\ref{fig:diff_nlo} that the final-state top quark and
Higgs boson momentum distributions become somewhat harder when we
increase the contributions from the $2\to3$ matrix elements, going to
$\mu_F=20 \GeV$ and $\mu_F=10 \GeV$. The same behavior is seen in the
invariant mass of the top quark and the Higgs boson pair, $m_{tH}$,
shown in Fig.~\ref{fig:diff_nlo}. To explore this feature we present a
normalized distribution of the $m_{tH}$ invariant mass, in which we
rescale the gluon distribution function for the central choice
$\mu_F=M/5$ by a factor $P_g(\mu_F,x_{\rm PDF})^2/P_g(M/5,x_{\rm
  PDF})^2$, to estimate the effect of the parton densities. Because
most of the cross section arises from production near threshold, we
can approximate $x_{\rm PDF}=m_{tH}/E_{\rm coll}$ with $E_{\rm
  coll}=14 \TeV$. The physics motivation of this cross check is that
the gluon densities become slightly harder for smaller scales, and we
want to understand whether the hardening of the $m_{tH}$ distribution
arises from the shift from the bottom parton picture to the gluon
fusion picture or if it is due to an overall hardening of the gluon
parton spectrum.  In Fig.~\ref{fig:diff_nlo} we observe that scaling
the usual NLO distribution with the $x$ dependence of the gluon parton
density reproduces the hardening of the top quark and Higgs boson
spectra.  Concluding this argument, we find a slight shift in the
spectrum at smaller factorization scales, but this shift is induced by
the shape of the gluon parton density. The two sets of distributions
in the bottom row of Fig.~\ref{fig:diff_nlo} show that there appears
to be no problem with the bottom parton approximation for sufficiently
inclusive distributions of the heavy final state particles at the LHC.
While there is a large logarithmic term $\log p_{T,b}/m_b$ present in
the $\bar{b}t\hm$ production rate, no additional
perturbative pitfalls appear in the $gb$ process after the
logarithmic terms are resummed. All additional dependence on the
(neglected) transverse momentum of the final-state bottom jet is
effectively power suppressed.\medskip

\subsection{Zero bottom quark mass approximation}

\begin{figure}[t]
 \begin{center}
 \includegraphics[width=15.0cm]{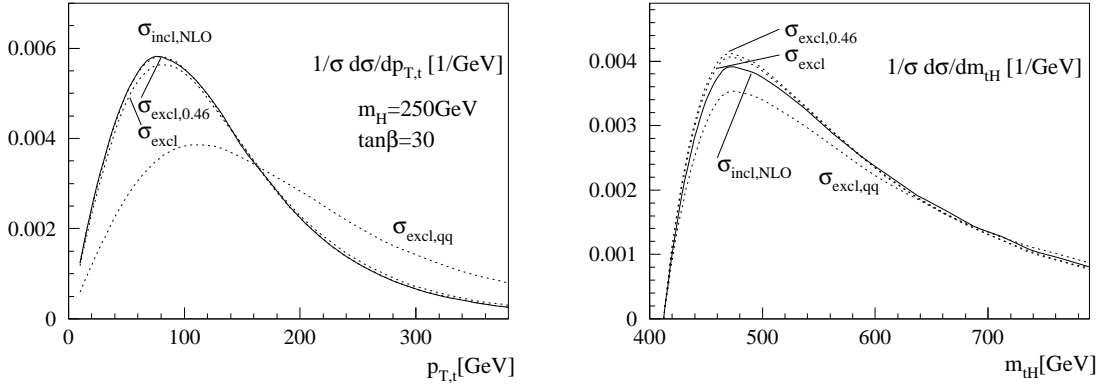}
 \end{center}
 \caption[]{The kinematic distributions for the final state at the
   LHC: the top quark transverse momentum and the invariant mass of
   the top quark and the Higgs boson pair. We compare the NLO result
   for the process $gb \to t\hm$ with the process
   $pp \to \bar{b}t\hm +X$.  Apart from the complete set of
   diagrams with the physical bottom quark mass, we show the purely
   $q\bar{q}$ induced process and the complete set of diagrams with a
   mathematical cutoff instead of the physical bottom quark mass $m_b
   \to 0.46\GeV$.} 
 \label{fig:diff_mb}
\end{figure}

We have compared the $gb$ induced process with the 
process $gg \to \bar{b}t\hm$ without a finite bottom quark mass in the
phase space or in the matrix element. For the kinematic distributions
of the bottom quarks this approximation is not obviously good. In our
NLO approach the divergences in the $p_{T,b}$ spectrum are compensated
by a negative infinity at $p_{T,b}=0$, \ie in the $2\to2$ kinematic
limit. This distribution is not physical, and all-orders soft gluon
resummation should be taken into account~\cite{resum} to obtain a
physical spectrum with a peak at some small value of $p_{T,b}$. With
the bottom quark mass as a regulator, the $p_{T,b}$ spectrum peaks
near $m_b$~\cite{old}. However, when the $gb$ process is
used, we are implicitly not interested in observing the final
bottom-quark jet and in its distributions; rather, we are interested
in the distributions of the heavy final state particles. In
Fig.~\ref{fig:diff_mb} we show the normalized transverse momentum
distribution of the top quark for the $gb$ process at NLO, and
for the $2 \to 3$ process, with two different cutoffs: one with the
physical bottom quark mass and the other with a smaller mathematical
cutoff (we use $1/10$ of the bottom quark pole mass). We observe that
the $gb$ calculation agrees with the $2 \to 3$ matrix element
approach with the physical bottom quark mass. The curve with a smaller
cutoff instead of the bottom quark mass agrees perhaps too well with
the NLO process in which the bottom quark mass is neglected.  The
distribution in the invariant mass of the $t\hm$ final state confirms
this level of agreement.  The dependence on the bottom quark mass
seems to be power suppressed, and the approximation of zero bottom
quark mass is justified.\smallskip

There are limitations of our argument for charged Higgs boson
production at the Tevatron. The $2 \to 3$ rate with a finite bottom
quark mass, induced by incoming quarks, shows considerably harder
momenta of final-state top quarks.  While the total cross section is
predicted correctly in the bottom parton picture, this effect might
mean that the kinematic distributions require more careful study at 
Tevatron energies.  The effect is not related to the approximation of 
zero bottom quark mass. Instead, it probes the perturbative link between
the contribution of gluon and quark initiated diagrams in the $2 \to 3$
production processes and the bottom parton
description.\medskip

Finally, we remark that the total cross section and the inclusive
distributions of the final state particles are correctly predicted in
the bottom parton picture.  Neither the small transverse momentum
approximation nor the small bottom quark mass approximation in the
bottom parton picture has a visible effect on the transverse momentum,
the rapidity, and the invariant mass distributions of the final state
top quark and Higgs boson.  Shifts induced by the massless bottom
parton approximation are washed out once detector resolution is taken
into account. For the processes under consideration, the contribution
from the $q\bar{q}$ initial states is less important at the LHC than
at the Tevatron.  The light-quark induced subprocesses show harder
transverse momentum spectra, a difference that should be considered
for predictions at Tevatron energies.

\subsection{Final state correlations}

The fully differential nature of the two--cutoff method enables us to
place a kinematic cut on one final state particle and to study the
distribution in the momentum of other particles in the final state.
It allows us to examine any correlation observable among the final
state particles which does not spoil the cancellation of soft and
(initial state) collinear divergences.  For reasons of acceptance
and/or to improve signal purity with respect to backgrounds, it may be
helpful to make final-state cuts that act in similar fashion to a cut
on the transverse momentum of the top-quark.  If the charged Higgs
boson decays to a $\tau$-lepton jet, a simple transverse momentum cut
on a lepton from the top quark decay could be an example. We examine
in this section how such a selection may affect the expected
transverse momentum distribution of the Higgs boson.  The Higgs boson
transverse momentum distribution is crucial because it determines the
boost of the Higgs boson decay products. If these products include
bottom-quark or $\tau$-lepton jets, their transverse momentum
distributions determine the tagging efficiencies.  \smallskip

\begin{figure}[t]
 \begin{center}
 \includegraphics[width=7.5cm]{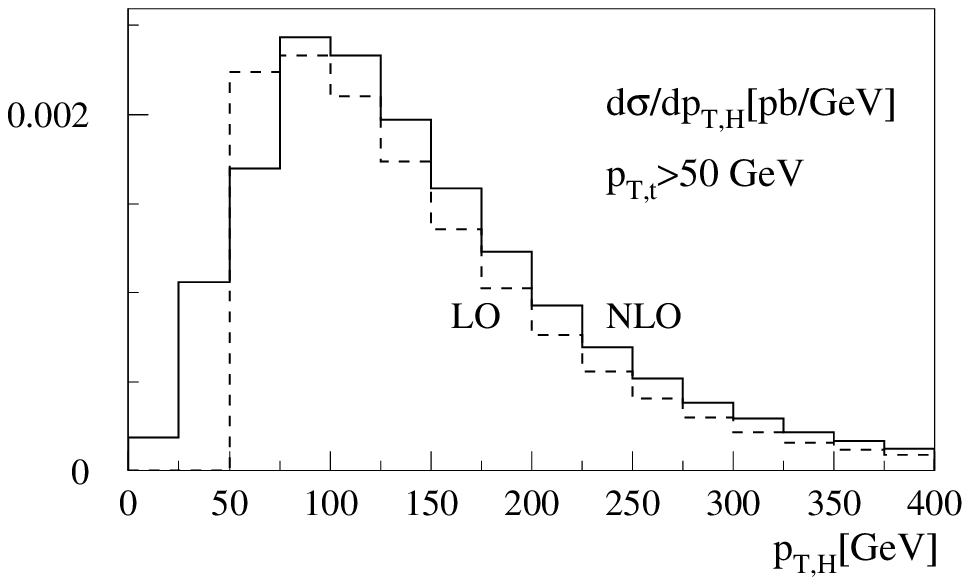} \hspace*{0.5cm}
 \includegraphics[width=7.5cm]{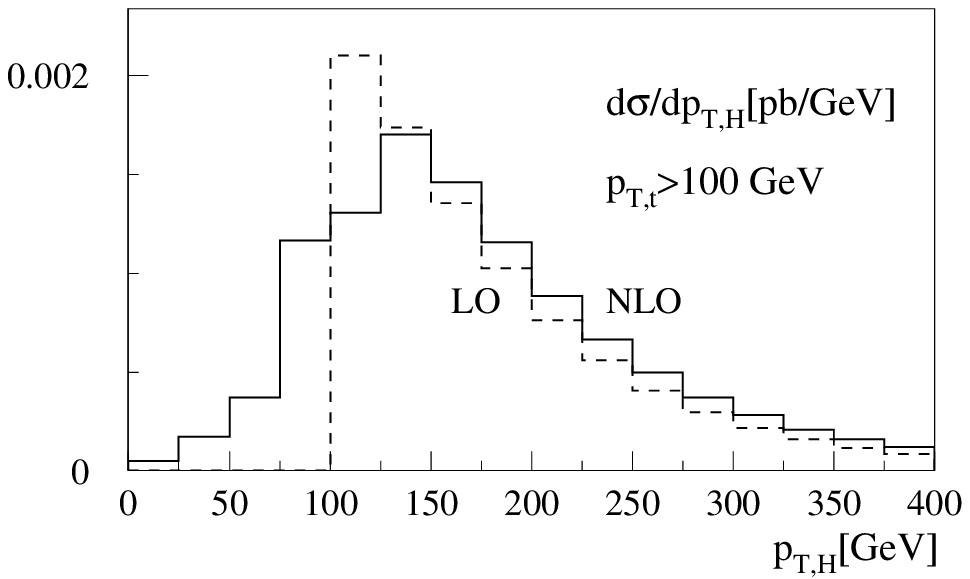}
 \end{center}
 \vspace*{-4mm}
 \caption[]{LO and NLO $p_T$-distribution of the Higgs boson, with
   cuts $p_{T,t}>50\GeV$ or $p_{T,t}> 100\GeV$ on the transverse
   momentum of the top quark. The Higgs boson mass is $m_H = 250\GeV$,
   and $\tan\beta = 30$.}
 \label{fig:pt_cut}
\end{figure}

The topology of the final state and the correlation in momentum
between the top-quark and the Higgs boson change once NLO corrections
are included. Instead of a back-to-back pair of heavy states at
leading order, the NLO topology includes three final-state particles
sharing the total transverse momentum.  In Fig.~\ref{fig:pt_cut}, we
show the $p_{T,H}$ distributions after the cuts $p_{T,t} > 50\GeV$ or
$p_{T,t} > 100\GeV$.  At LO, a cut on $p_{T,t}$ indeed eliminates
values of $p_{T,H}$ below this cut. The LO distribution
$d\sigma/dp_{T,H}$ in the range $p_{T,H} > 100\GeV$ is identical for
the two cases shown: $p^{\rm min}_{T,t} = 50\GeV$ and $p^{\rm
  min}_{T,t} = 100\GeV$.  At NLO, the figure shows that the effects of
cuts on the momentum of one final state particle extend over a
significant range in the momentum of another final state particle.
Owing to the presence of the final state jet, the NLO transverse
momentum distribution of the Higgs boson extends all the way to zero,
as shown in Fig.~\ref{fig:pt_cut}.  The impact of the $p_{T,t}$ cut on
the NLO distributions is evident well above $p_{T,H} = p^{\rm
  min}_{T,t}$, and the distributions do not coincide until $p_{T,H} >
200\GeV$.  The ratio of the NLO and LO distributions cannot be
represented by a simple correction (or ``$K$'') factor. This factor
would be infinite for values of $p_{T,H}$ less than the value of the
cut on $p_{T,t}$, less than unity in a small interval where $p_{t,H}$
is just above the cut, and uniformly greater than unity for $p_{T,H} >
1.5 p_{T,t}^{\rm min}$.

\begin{figure}[b]
 \begin{center}
 \includegraphics[width=6.5cm]{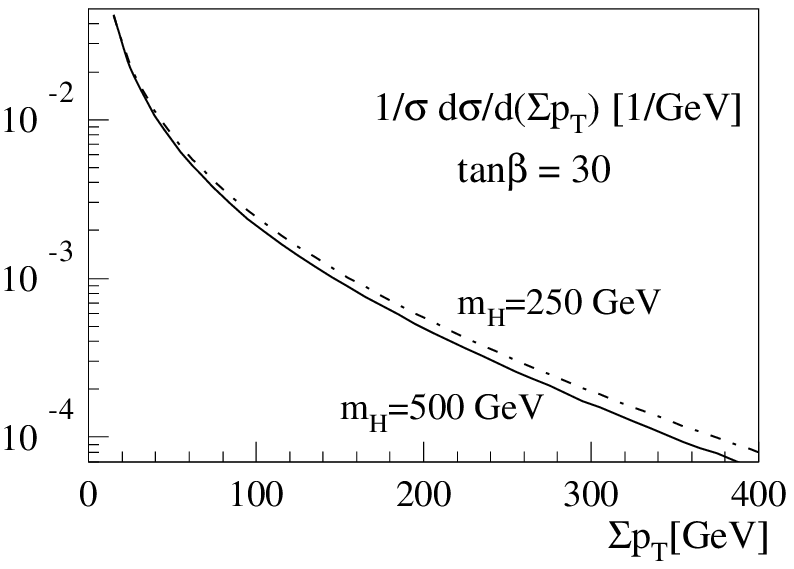} \hspace*{1.0cm}
 \includegraphics[width=7.0cm]{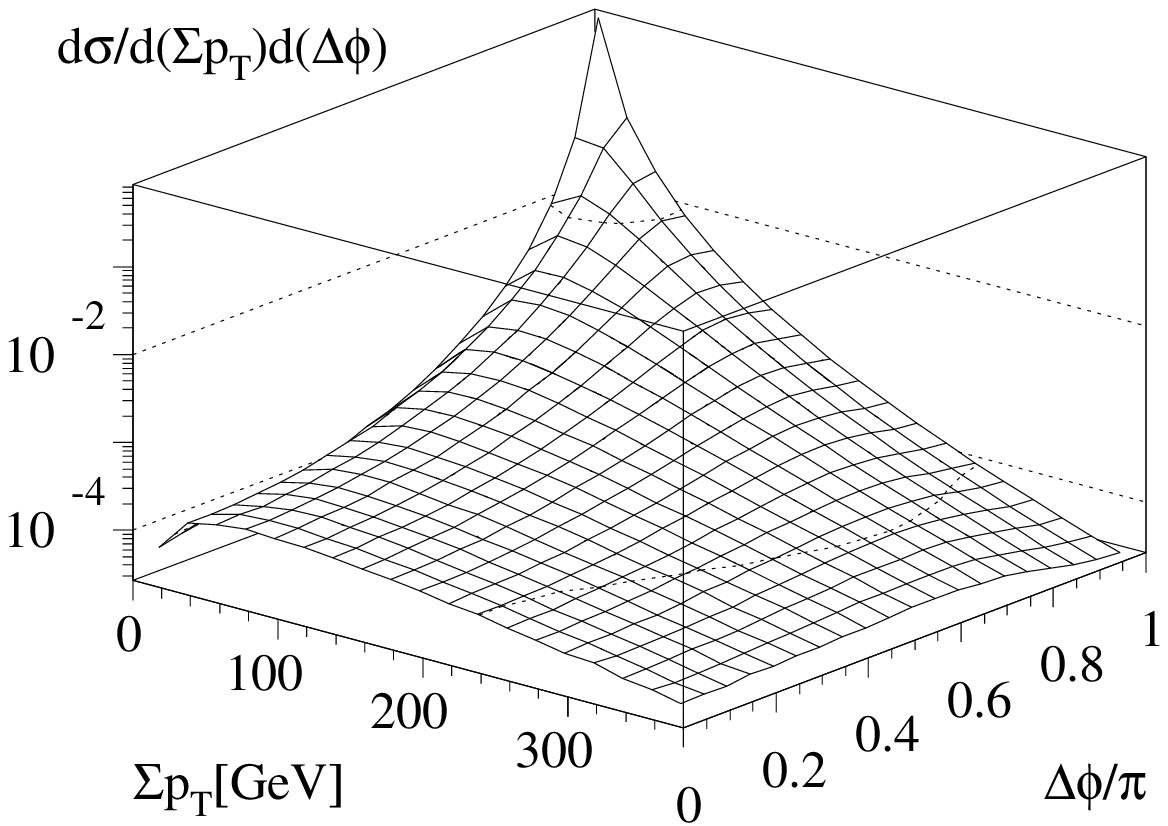}
 \end{center}
 \vspace*{-4mm}
 \caption[]{Left: Normalized distributions in $\Sigma p_T$ (defined 
   in the text) for
   the NLO process $gb \to tH^-$, for $m_H = 250\GeV$ and
   $500\GeV$.  Right: A two-dimensional plot of the correlation 
   between $\Sigma p_T$ and $\Delta\phi$, for $m_H = 250\GeV$ and 
   $\tan\beta = 30$.}
 \label{fig:spdphi}
\end{figure}

In contrast to the differences seen in Fig.~\ref{fig:pt_cut}, the
effect of a $p_{T,t}$ cut on the Higgs boson rapidity distribution is
trivial: the NLO rapidity distributions of the Higgs boson and the
top-quark are reduced by an approximately uniform factor.

The distribution in $p_{T,H}$ is well defined at NLO.  The soft and
the initial-state collinear divergences appear with Born--type
kinematics, just like the explicit infrared poles.  However, we expect
strong cancellations between large NLO negative virtual contributions
and positive NLO real emission contributions at the LO threshold.
Numerical problems can arise once these cancellations yield a cross
section at threshold that is more than an order of magnitude smaller
than the contributing parts.  Motivated by evidence of instability 
in the threshold region $p_{T,H} = p^{\rm cut}_{T,t}$ if we use the 
values of $\delta_s$ and $\delta_c$ chosen earlier, we select cutoff 
values $\delta_s = 10^{-2}$ and $\delta_c = 2 \times 10^{-4}$ to obtain 
the results shown in Fig.~\ref{fig:pt_cut}.  These relatively large 
values of the cutoffs remain in the safe region for the total cross 
section, as shown in Fig.~\ref{fig:cutoff}. Moreover, we use fairly 
wide bins, to be less sensitive 
to numerical cancellations in the threshold region. The behavior
of the curves in Fig.~\ref{fig:pt_cut} near threshold indicates a 
some remaining uncertainty, so the $p_{T,H}$ 
distributions should be taken with reservation at threshold.  
They should be reliable one bin or more from threshold in each 
direction.\medskip
 
Another distribution of interest is the vector sum of the transverse
momenta of the top quark and the Higgs boson ($\Sigma p_T =
|(\vec{p}_{T,t} + \vec{p}_{T,H})|$) shown in the left panel of
Fig.~\ref{fig:spdphi}. The LO distribution is a single peak at $\Sigma
p_T = 0$ since the top quark and the Higgs boson balance in transverse
momentum at this order.  At NLO, the observable $\Sigma p_T$ is the
transverse momentum of the third jet in the final state.  The shape of
the distribution depends only mildly on the Higgs boson mass.  The NLO
distribution in $\Sigma p_T$ shows a marked divergence as $\Sigma p_T
\rightarrow 0$.  This divergence reflects a limitation of our
fixed-order calculation and points to the eventual need for all-orders
resummation of the effects of soft-gluon radiation, as discussed and
implemented for other processes~\cite{resum}.
   
The right panel of Fig.~\ref{fig:spdphi} displays the two-dimensional
correlation between $\Sigma p_T$ and ${\Delta \phi = | \phi_t -
  \phi_H|}$ for the sum of all $2 \to 3$ contributions to the cross
section.  Here $\Delta \phi$ is the difference between the azimuthal
angles of the Higgs boson and the top quark.  At LO it is fixed to
$\Delta \phi = \pi$, but for the $2\to 3$ processes the distribution
extends to all $\Delta \phi$, again a manifestation of the fact that
an additional parton in the final state contributes to the transverse
momentum balance. It is interesting to see that the $\Sigma p_T$
distribution for $\Delta \phi \ne \pi$, \ie for a topology different
from LO, develops a maximum for $\Sigma p_T \sim 50\GeV$ and drops to
zero for $\Sigma p_T \to 0$. This feature and the corresponding
behavior of $\Delta \phi$ for $\Sigma p_T \ne 0$ reflect the fact that
the (resummable) divergence is limited to the LO topology.

\section{Supersymmetric contributions}
\label{sec:susy}

\begin{table}[t]
\begin{center}
{\scriptsize
\begin{tabular}{|c||cc|ccccc|c||c|c||r|r|r|}
\hline
mSUGRA & $m_H$ & $\tan\beta$ & $m_0$ & $m_{1/2}$ & $A_0$ & & & $\mu$ & $\sigma_{\rm LO}$[fb] & $\sigma_{\rm NLO}$[fb] & $\Delta_b$ & $\Delta_b^{\rm resum}$ & non--$\Delta_b$ \\
\hline
1a & 402  & 10 & 100  & 250 &   -100 & & & 352 & 18.7  & 25.6  & -11.0\% & -10.2\% &  -1.9\% \\
1b & 543  & 30 & 200  & 400 &      0 & & & 501 & 47.1  & 61.7  & -27.9\% & -23.5\% &  -4.6\% \\
2  & 1446 & 10 & 1450 & 300 &      0 & & & 125 & 0.09  & 0.13  & -0.92\% & -0.91\% &  -1.7\% \\
3  & 578  & 10 &   90 & 400 &      0 & & & 509 & 5.81  & 8.02  & -10.1\% &  -9.5\% &  -1.1\% \\
4  & 416  & 50 &  400 & 300 &      0 & & & 377 & 304 & 395 & -39.0\% & -31.0\% &  -4.6\% \\
5  & 699  &  5 &  150 & 300 &  -1000 & & & 640 & 3.73  & 5.73  & -8.5\% &  -8.0\% &  0.8\% \\
\hline
mSUGRA-like &  & & $m_0$ & $m_{1/2}$ & $A_0$ & 
               $M_1$ & $M_{2,3}$ & & & & & & \\
\hline      
6  & 470 & 10 & 150 & 300 &      0 & 480 & 300 & 394 & 11.6 & 16.0 & -10.2\% & -9.5\% & -1.3\% \\ 
\hline
GMSB & & & $\Lambda$ & $M_{\rm mes}$ & $N_{\rm mes}$ &
 & & & & & & & \\ 
\hline     
7  & 387 & 15 & $ 40 \times 10^3$ & $ 80 \times 10^3$ &      3 & & & 300 & 36.5 & 48.0 & -8.5\% & -8.1\% &  -0.9\% \\    
8  & 521 & 15 & $100 \times 10^3$ & $200 \times 10^3$ &      1 & & & 398 & 15.0 & 20.4 & -7.5\% & -7.1\% &  -0.5\% \\   
\hline
AMSB & & & $m_0$ & $m_{\rm aux}$ & & & & & & & & & \\ 
\hline         
9  &  916 & 10 & 400 &  $60 \times 10^3$ &        & & & 870 & 0.92 & 1.29 & -10.6\% & -9.9\% &  4.1\% \\
\hline
\end{tabular}
}
\end{center}
\caption[]{Supersymmetric corrections to the production cross section
  $gb \to t\hm$ from non-resummed and resummed $\Delta_b$ corrections,
  Eq.~(\ref{eq:delta_mb}), and from the explicit remaining
  supersymmetric loop diagrams. The supersymmetric parameter points
  are chosen according to the benchmarks in Ref.~\cite{sps}.  All
  masses are given in units of GeV. The percentage changes are defined
  with respect to the purely gluonic NLO rates.}
\label{tab:sps}
\end{table}

Supersymmetric diagrams contribute to the production rate for $gb \to
t\hm$ at the same level as the NLO QCD contributions ($\alpha_s^2
y_{b,t}^2$), These diagrams are virtual gluon exchange diagrams, where
the gluons and quarks are replaced by their supersymmetric partners,
gluinos and squarks.  A feature of squarks is the mixing between the
supersymmetric partners of the left-handed and the right-handed
quarks. The $2\times2$ bottom-squark mass matrix has an off-diagonal
entry $m_b (A_b - \mu\tan\beta)$, and the top-squark mass matrix has
an entry $m_t (A_t - \mu/\tan\beta)$.  Here $\mu$ is the Higgsino mass
parameter which links the two Higgs doublets in the Lagrangian, and
$A_i$ is the trilinear squark-squark-Higgs boson coupling parameter.
The off--diagonal entry in the matrix element is proportional to the
mass of the standard model partner, and it is usually neglected for
the first and second generations. For large values of $\tan\beta$
bottom squark mixing can become larger than top squark mixing.

In the limit of large $\tan\beta$, the leading supersymmetric
contributions are not loop diagrams, but renormalization
terms~\cite{delta_mb}. This point becomes evident if we compute the
corrections to the $bt\hm$ vertex in the limit of vanishing bottom
quark mass, but finite bottom quark Yukawa coupling. This approach is
justified from a formal point of view because the connection between
the mass and the Yukawa coupling is a property of electroweak
symmetry-breaking and not protected once the symmetry is broken.
Disassociation of the mass and the Yukawa coupling becomes apparent in
a type-II two-Higgs-doublet model like the MSSM.  Because the bottom
quark Yukawa coupling always appears as $m_b \tan\beta$, the
relationship between the mass and the Yukawa coupling is not fixed. If
we compute the renormalization of the $tb\hm$ vertex with a zero
external bottom quark mass, bottom squark mixing diagrams lead to
contributions which look like mass renormalization terms, \ie terms
which create a finite bottom quark mass in the external leg. However,
this interpretation cannot be correct, since mass renormalization has
to be multiplicative. Instead, we see that these renormalization
factors describe a misalignment of the bottom quark Yukawa coupling
and the bottom quark mass, which appears for zero bottom quark mass as
well as for finite values. In complete analogy to a mass
renormalization, these coupling renormalization diagrams can be
resummed to all orders and lead to a correction~\cite{delta_mb_resum}:
\begin{alignat}{7}
\frac{m_b \tan\beta}{v} \; &\to \; \frac{m_b \tan\beta}{v} \;
                           \frac{1}{1+\Delta_b} \notag \\ \Delta_b &=
                           \; \frac{\sin(2 \theta_b)}{m_b} \;
                           \frac{\alpha_s}{4 \pi} \; C_F \;
                           m_{\tilde{g}} \; \frac{1}{i \pi^2} \;
                           \left[ B(0,m_{\tilde{b},2},m_{\tilde{g}})
                           -B(0,m_{\tilde{b},1},m_{\tilde{g}}) \right]
                           \phantom{haaallllooooooo} \notag \\ &= \;
                           \frac{\alpha_s}{2 \pi} \; C_F \;
                           m_{\tilde{g}} \; \left(- A_b + \mu \tan\beta
                           \right) \;
                           I(m_{\tilde{b},1},m_{\tilde{b},2},m_{\tilde{g}})
                           \notag \\[3mm] I(a,b,c) &= -
                           \frac{1}{(a^2-b^2)(b^2-c^2)(c^2-a^2)} \;
                           \left[ a^2b^2 \log \frac{a^2}{b^2} +b^2c^2
                           \log \frac{b^2}{c^2} +c^2a^2 \log
                           \frac{c^2}{a^2} \right] .
\label{eq:delta_mb}
\end{alignat}
The function $B(p^2,m_1,m_2)$ is the usual scalar two--point function;
$C_F=4/3$ is the Casimir factor in the fundamental representation of
$SU(3)$. There are similar additional terms proportional to the strong
coupling or to the top quark Yukawa coupling, but
Eq.~(\ref{eq:delta_mb}) is the leading contribution for large
$\tan\beta$.

\begin{figure}[t]
 \begin{center}
 \includegraphics[width=15.0cm]{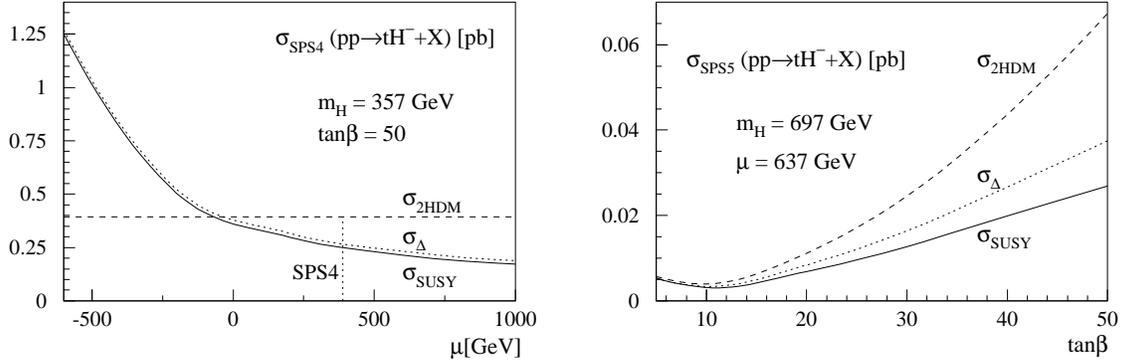}
 \end{center}
 \caption[]{The supersymmetric contributions to the NLO 
   cross section for $gb \to t\hm$, with the dotted curve
   showing the resummed $\Delta_b$ contribution and the solid curve
   showing the combination of the resummed $\Delta_b$ and
   non-$\Delta_b$ contributions. The supersymmetric parameters are
   described in Table \ref{tab:sps}. The central value of $\mu$ in
   SPS4 is noted in the plot; the central value of $\tan\beta$ in SPS5
   is at the lower end of the plot, $\tan\beta=5$. All parameters and
   masses except for $\mu$ and $\tan\beta$ are kept constant.  The
   curves denoted $\sigma_{\rm 2HDM}$ show NLO QCD cross sections
   without SUSY contributions.}
 \label{fig:sps}
\end{figure}

Since these $\Delta_b$ corrections are the leading
$\tan\beta$-enhanced supersymmetric contributions to the production
cross section, and since the charged Higgs boson search is most
promising at $\tan\beta \gtrsim 15$, we might speculate that these
corrections to the $gb \to t\hm$ production rate are sufficient.
Equation~(\ref{eq:delta_mb}) shows that the shift in the Yukawa
coupling can have large effects~\cite{sola} provided that $\mu
\tan\beta$ is large (preferably negative), the gluino mass is large,
and bottom-squark masses are not too large.  In this limit the
percentage corrections are approximately $\Delta_b \sim \mu
\tan\beta/m_{\tilde{g}}$.  The non-$\Delta_b$-type supersymmetric
corrections are negligible compared, for example, to the remaining NLO
scale variation. \medskip

The supersymmetric contributions to the charged Higgs boson production
cross sections are shown in Table~\ref{tab:sps} for the ``Snowmass
points and slopes'' (SPS)~\cite{sps} parameters. The contributions are
split into the $\Delta_b$ corrections, as defined in
Eq.~(\ref{eq:delta_mb}), and the remaining supersymmetric diagrams.
We present the $\Delta_b$ contributions in the NLO version
$1-2\Delta_b$ as well as after resummation, $1/(1+\Delta_b)^2$.  The
ratios of the SUSY corrections to the NLO QCD cross sections are given
in the last three columns. The negative sign of the $\Delta_b$
contributions is fixed by the sign of $\mu$. The sign of $\mu$ is a
free parameter, linked to SUSY contributions to the transition rate
$b\to s\gamma$.  In this process the measured rate is consistent with
the standard model prediction.  There are additional charged Higgs
boson and chargino induced contributions in the MSSM.  For $\mu>0$
they enter with opposite signs and therefore tend to cancel
numerically, while for $\mu<0$, in particular in the mSUGRA
supersymmetry breaking scheme, the parameter space is closely
constrained. Therefore, all SPS points are chosen with positive sign
of $\mu$.  \smallskip

In Table~\ref{tab:sps} we observe that the $\Delta_b$ corrections are
dominant for all points with $\tan\beta \ge 15$, particularly for the
two points with $\tan\beta=30,50$.  The leading contribution for large
$\tan\beta$ is described correctly by the $\Delta_b$ corrections.  At
maximum, all supersymmetric corrections are of the order of the
remaining scale variation and our estimate of the theoretical
uncertainty of $20 \%$, as long as $\tan\beta \lesssim 30$. This
modest correction is not necessarily true for the entire
supersymmetric parameter space, and the $\Delta_b$ corrections can be
much larger~\cite{sola}. However, the small correction reflects the
ansatz used in supersymmetry breaking. None of the scenarios in
Table~\ref{tab:sps} is designed to produce a large splitting in the
supersymmetric mass parameters at the weak scale or a large value of
$|\mu|$, which would favor large $\Delta_b$-type corrections.  In
general, large values of $|\mu|$ are a challenge in high-scale
motivated models.  For fine-tuning reasons these models usually
produce $|\mu|$ of order the weak scale, to avoid large cancellations
of different renormalization group contributions to the value of
$m_Z$.  Even in the focus-point~\cite{focus} inspired SPS2 the
contribution to weak--scale parameters that are proportional to $m_0$
cancels in itself, decoupling the value of $m_0$ from the leading
renormalization group running.  All other parameters remain at typical
weak-scale values.  In all three SUSY breaking scenarios considered,
the large gluino mass, linked to the relative dominance of the
corresponding beta function $\beta_3$, is the reason the $\Delta_b$
correction is not negligible.\medskip

\begin{figure}[t]
 \begin{center}
 \includegraphics[width=15.0cm]{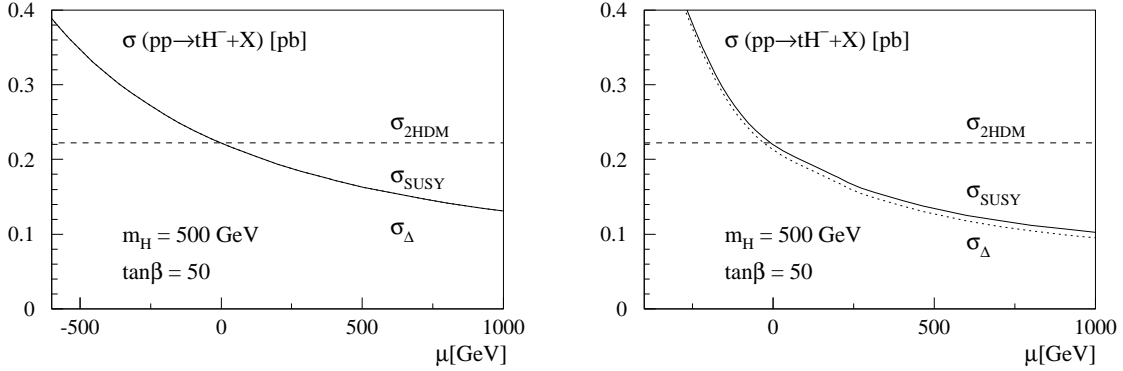}
 \end{center}
 \caption[]{The supersymmetric contributions to the NLO prediction for
   the cross section $gb \to t\hm$. The supersymmetric
   parameter points are the scenarios A and C, picked from
   Ref.~\cite{sola}. All parameters and masses except for $\mu$ are
   kept constant. The gluino mass in both scenarios is $1\TeV$; the
   lighter top-squark and bottom-squark masses are $0.5\TeV$ for
   scenario A (LHS) and $1\TeV$ for scenario C (RHS). The top-squark
   mass difference is $100\GeV$ and the bottom-squark mass difference
   $150\GeV$.  In contrast to Ref.~\cite{sola} we resum only the
   SUSY-QCD corrections.  The curves denoted $\sigma_{\rm 2HDM}$ show
   NLO QCD cross sections without SUSY contributions.}
 \label{fig:bggs}
\end{figure}

In Fig.~\ref{fig:sps} we show that the $\Delta_b$ contributions can
become large once we depart from the unification scenarios.  Starting
from the mSUGRA motivated points SPS4 and SPS5, listed in
Table~\ref{tab:sps}, we vary $\mu$ and $\tan\beta$, leaving all other
masses and parameters invariant. As expected, the non-resummed
$\Delta_b$ corrections become arbitrarily large for large values of
$|\mu|$, and the resummed $\Delta_b$ correction can become arbitrarily
large for some negative values of $\mu$, both limited only
perturbatively and ultimately by unitarity of the enhanced bottom
quark Yukawa coupling.  The sign of the $\Delta_b$ correction is fixed
by the sign of $\mu$, and the remaining SUSY contributions are small.
For comparison, we also show the NLO cross sections without SUSY
contributions included, the typical case for a two-Higgs-doublet model
(2HDM).  These curves are labeled $\sigma_{\rm 2HDM}$.  The NLO QCD
cross section does not depend on the SUSY parameter $\mu$, but it
increases as $\tan\beta$ gets large, as expected.
In Fig.~\ref{fig:bggs} we show the same effect, starting from the
scenarios A and C in Ref.~\cite{sola}. Because the value of
$\tan\beta=50$ is large, the non--$\Delta_b$ corrections are
completely negligible, while the $\Delta_b$ corrections can become
arbitrarily large. The difference in the size of the corrections in
the two panels of Fig.~\ref{fig:bggs} can be understood from
Eq.~(\ref{eq:delta_mb}) in the limit $a \sim b$ for the bottom squark
masses and either $c \gg a,b$ or $c\sim a,b$.  In both cases the
$\Delta_b$ corrections are suppressed by the heaviest mass in the
system, \ie the gluino mass, but the pre-factor is larger if all
masses involved are of the same order.

\section{Conclusions}
\label{sec:con}

We evaluate the inclusive and differential cross sections for the
associated production of a top quark along with a charged Higgs boson
at Tevatron and LHC energies to next-to-leading order in QCD and in
supersymmetric QCD. \medskip

Using the two--cutoff scheme to treat the soft and collinear
singularities, we find stable results for total and differential cross
sections over large ranges of the cutoff parameters as well as of the
factorization and renormalization scales. While the QCD corrections to
the total rate are sizable at the LHC, $K\sim 1.4$~\cite{old}, the
shifts in the normalized kinematic distributions of the heavy final
state top quark and Higgs boson are negligible. The scale dependence
gives us a reasonable estimate of about $20\%$ on the remaining
theoretical uncertainty. \medskip

In the regime where $m_H < m_t$, we compute the NLO cross section by
subtracting the intermediate on-shell divergences in the narrow width
approximation. This procedure allows us to match the NLO cross section
for the process $gb \to t\hm$ with the contributions from $gg \to t
\bar{t}^*$ with a subsequent decay $\bar{t}^* \to \bar{b} \hm$, simply
by adding the rates.  This method yields a prediction for the cross
section for associated charged Higgs boson production over the entire
range of Higgs boson.\medskip

At the Tevatron, charged Higgs boson production is likely to be
observed only for small masses of the Higgs boson. In this regime we
show the validity of the Breit--Wigner approximation in the $t\bar{t}$
production process.  We can add the off-shell production rate at
NLO.\medskip

Examining the NLO momentum distributions for inclusive charged Higgs
boson production, we show the validity of the bottom parton
description beyond the total rate. Neither the collinear phase space
approximation nor the approximation of zero bottom quark mass has a
visible impact on the kinematic distributions of the heavy final state
particles.\medskip

The fully differential nature of the two--cutoff method enables us to
place a kinematic cut on one final state particle and study the
distribution in momentum of the other particles.  It allows us also to
examine momentum correlations among the final state particles.
\medskip

We explore the effects of virtual supersymmetric particles in NLO loop
diagrams and find that the universal $\Delta_b$ corrections to the
Yukawa coupling can be sizable. In the two Higgs doublet model, the
remaining explicit loop contributions to the NLO rate are below the
level of the scale uncertainty.

\acknowledgments We thank Brian Harris for his assistance with the
two--cutoff method and Shouhua Zhu for providing comparisons between
our results and those in Ref.~\cite{zhu}.  We are grateful for helpful
discussions with Carlos Wagner and John Campbell.  J.J.  acknowledges
helpful conversations with Jungil Lee.  T.P. wants to thank Sasha
Nikitenko for his support and for pointing out questions from the
experimental community. Moreover, T.P. wishes to thank Thomas Teubner
and Sakis Dedes for helpful discussions, and last but not least the
IPPP in Durham for their kind hospitality during the final stage of
this paper.  The research of E.~L.~B. and J.~J.~in the High Energy
Physics Division at Argonne National Laboratory is supported by the
U.~S.~Department of Energy, Division of High Energy Physics, under
Contract W-31-109-ENG-38.  T.H. is supported in part by the DOE under
grant DE-FG02-95ER40896, and in part by the Wisconsin Alumni Research
Foundation.

\end{document}